\documentclass[
    a4paper,
    twocolumn,
    10pt,
    accepted=2026-02-21,
]{quantumarticle}
\pdfoutput=1

\usepackage[utf8]{inputenc}
\usepackage[english]{babel}
\usepackage[T1]{fontenc}

\usepackage{amsmath,amssymb}
\usepackage{graphicx}
\usepackage{dcolumn}
\usepackage{bm}
\usepackage{bbold}
\usepackage{varioref}
\usepackage{quantikz}
\usepackage{tikz}
\usepackage{tikz-cd}
\usetikzlibrary{quantikz2}
\usepackage{braket}
\usepackage{soul}
\usepackage{tabularx}
\usepackage[export]{adjustbox}

\usepackage[numbers]{natbib}

\newcommand{\pt}[1]{\left( #1 \right)} 
\newcommand{\pq}[1]{\left[ #1 \right]}
\newcommand{\pg}[1]{\left\{ #1 \right\}}
\newcommand{\pa}[1]{\left\langle #1 \right\rangle}

\newcommand{\me}{\mathrm{e}} 
\newcommand{\ramuno}{\mathrm{i}} 
\newcommand{\id}{\mathbb{1}} 
\DeclareMathOperator{\Tr}{Tr} 
\newcommand{\transpose}{\intercal} 
\newcommand{\imag}{\mathfrak{Im}} 
\newcommand*\de{\mathop{}\!d} 

\newcommand{\tham}{H} 
\newcommand{\uham}{H_0} 
\newcommand{\source}{H_\mathrm{s}} 
\newcommand{\gs}{\ket{\psi_0}} 
\newcommand{\cgs}{\bra{\psi_0}} 
\newcommand{\fdfunc}{F} 
\newcommand{\nperturb}{N} 
\newcommand{\nqubits}{L} 

\newcommand{\var}{\mathrm{var}} 
\newcommand{\eval}{\mathbb{E}} 
\newcommand{\grad}{\hat{g}^{\pt{k}}} 
\newcommand{\efunc}{E} 
\newcommand{\exfunc}{\hat{E}} 

\makeatletter
\AddToHook{cmd/appendix/before}{\def\cref@section@alias{appendix}\def\cref@subsection@alias{appendix}}
\makeatother

\definecolor{blue}{RGB}{68,119,170}
\definecolor{green}{RGB}{34,136,51}
\definecolor{qviolet}{RGB}{83,37,127}
\definecolor{qgray}{RGB}{85,85,85}

\usepackage[
colorlinks=true,
citecolor=qviolet,
urlcolor=qviolet,
linkcolor=qviolet,
]{hyperref}
\usepackage[capitalise]{cleveref}
\usepackage[acronym]{glossaries}
\glsdisablehyper

\newacronym
  [plural=RGFs]
  {rgf}
  {RGF}
  {Retarded Green's Function}
\newacronym
  {lr}
  {LR}
  {Linear Response}
\newacronym
  {dsf}
  {DSF}
  {Dynamical Structure Factor}
\newacronym
  {fd}
  {FD}
  {Finite Differences}
\newacronym
  {psr}
  {PSR}
  {Parameter-Shift Rule}
\newacronym
  {spsa}
  {SPSA}
  {Simultaneous Perturbation Stochastic Approximation}
\newacronym
  {lcp}
  {LCP}
  {Local Circuit Perturbation}
\newacronym
  {scp}
  {SCP}
  {Simultaneous Circuit Perturbation}
\newacronym
  {jw}
  {JW}
  {Jordan--Wigner Transformation}
\newacronym
  {bic}
  {BIC}
  {Bayesian Information Criterion}

\newcommand{\affiliationibm}{IBM Research Europe -- Zurich, S\"aumerstrasse 4, 8803 R\"uschlikon, Switzerland}
\newcommand{\affiliationepfl}{Institute of Physics, \'Ecole Polytechnique F\'ed\'erale de Lausanne (EPFL), CH-1015 Lausanne, Switzerland}

\begin{document}

\title{A circuit-differentiation framework for Green's functions on quantum computers}

\author{Samuele Piccinelli}%
\email{samuele.piccinelli@ibm.com}
\affiliation{%
\affiliationibm
}%
\affiliation{%
\affiliationepfl
}%
\orcid{0000-0003-3782-7565}%
\author{Francesco Tacchino}%
\email{fta@zurich.ibm.com}
\affiliation{%
\affiliationibm
}%
\orcid{0000-0003-2008-5956}%
\author{Ivano Tavernelli}%
\affiliation{%
\affiliationibm
}%
\orcid{0000-0001-5690-1981}%
\author{Giuseppe Carleo}%
\affiliation{%
\affiliationepfl
}%
\orcid{0000-0002-8887-4356}%


\maketitle

\begin{abstract}
We propose a general framework for computing \glspl{rgf} on quantum computers by recasting their evaluation as a problem of circuit differentiation. Our proposal is based on real-time evolution and specifically designed circuit components, which we refer to as circuit perturbations, acting as a direct representation of the external perturbative force within the quantum circuit in a linear-response setting. The direct mapping between circuit derivatives and the computation of \glspl{rgf} enables the use of a broad range of differentiation strategies. We provide two such examples, including a class of stochastic estimators which do not require extra qubit connectivity with respect to the underlying time-evolution operations. We demonstrate our approach on interacting spin and fermionic models, showing that accurate dynamical correlations can be obtained even under realistic noise assumptions. Finally, we outline how our proposal can be tied to efficient gradient-estimation techniques relevant for the fault-tolerant regime.
\end{abstract}

\glsresetall

\section{\label{sec:introduction}Introduction}

Understanding the dynamical properties of quantum many-body systems is a central question in virtually all areas of modern physical sciences, including quantum chemistry, condensed matter and high-energy physics. In particular, the study of low-energy excitations often provides a direct bridge between theoretical models and experimental practice. However, because of the strongly correlated nature of many relevant models and the associated rapid growth of entanglement during time evolution, their investigation remains challenging at large scale. In the last decade, quantum computers have emerged as a promising computational paradigm to address this class of problems~\cite{miessen_quantum_2022,di_meglio_quantum_2024,alexeev_quantum-centric_2024}, and recent demonstrations of quantum simulations at the so-called quantum utility scale~\cite{kim_evidence_2023} have opened the way to the use of digital quantum information processing platforms as concrete research tools for the natural sciences.

In this context, time correlation functions of the form
\begin{equation}
    \mathcal{C}\pt{t-t_0}=\Tr\pq{\rho A\pt{t}B\pt{t_0}}
\end{equation}
with $t>0$ are particularly useful to characterize the behavior of the systems under study. In the equation above, $\rho$ is a reference many-body state and $A$, $B$ are operators associated with a particular physical process. These time correlation functions are intimately connected to the \glspl{rgf}, which connect theoretical predictions with experimental observables, ranging from dynamical modes and spectral functions to scattering amplitudes. The ability to compare theoretical predictions based on these functions against techniques like angle-resolved photoemission or inelastic neutron scattering constitutes a powerful tool in the hands of computational quantum scientists. Nevertheless, computing dynamical correlation functions in large-scale quantum systems remains a significant challenge for classical simulation techniques.
\begin{figure}
\includegraphics{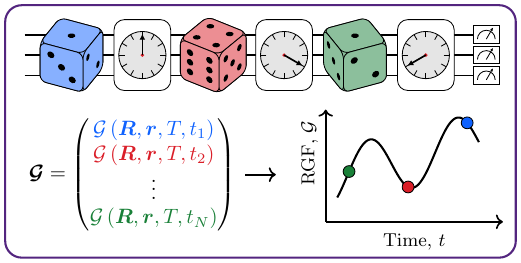}
\caption{\textbf{Circuit perturbation method.} Our approach builds on results from linear response theory and exploits a correspondence between quantum circuits derivatives and generalized susceptibilities. By inserting randomized perturbations throughout the system evolution and leveraging stochastic gradient-estimation tools, we obtain information about \glspl{rgf} at different time intervals in parallel from a single quantum circuit template.}\label{fig:abstract}
\end{figure}
For these reasons, quantum algorithms to compute \glspl{rgf} in both the time and frequency domains have been extensively explored in the literature~\cite{somma_simulating_2002,tazhigulov_simulating_2022,pedernales_efficient_2014,roggero_linear_2019}. Examples include quantum subspace expansion algorithms~\cite{jamet_quantum_2022}, a generalization of the quantum Equation-of-Motion (qEOM) protocol leveraging the computation of excited states~\cite{ollitrault_quantum_2020,rizzo_one-particle_2022,selisko_dynamical_2024}, variations of the Lanczos algorithm~\cite{baker_lanczos_2021,jamet_krylov_2021} and the continued fraction representations of dynamical correlation functions~\cite{irmejs_approximating_2025}. In the time domain, a standard approach is to use a suitable version of the well-known ancilla-based Hadamard test~\cite{chiesa_quantum_2019,tacchino2020quantum}, which requires long-range controlled quantum operations. While measurements of out-of-time-ordered correlators have been demonstrated on analog quantum simulators~\cite{garttner2017measuring}, large-scale simulations on digital platforms have remained limited, particularly on architectures with restricted qubit-qubit connectivity. Proof-of-principle demonstrations on digital devices include examples based on real-time dynamics~\cite{baroni_nuclear_2022,chiesa_quantum_2019,braumuller_probing_2022}, imaginary-time dynamics~\cite{tazhigulov_simulating_2022,sun_quantum_2021} and variational approaches~\cite{endo_calculation_2020,libbi_effective_2022,huang_variational_2022}.

Alternative strategies to access dynamical susceptibilities without requiring ancilla qubits have been explored~\cite{mitarai_methodology_2019,kastner_ancilla-free_2024}, and, in particular, methods inspired by \gls{lr} experiments have been recently proposed~\cite{baez_dynamical_2020,kokcu_linear_2024,Bishop}. Building on these ideas, here we present a comprehensive framework for computing \glspl{rgf} on quantum processors, and construct a novel class of quantum algorithms which are already suitable for experiments on noisy devices. More specifically, we develop and extend the theoretical analysis of previous studies to make explicit the connection between \gls{lr} functions and derivatives of parametrized quantum circuits. Building on this formulation, we introduce circuit constructions that allow \glspl{rgf} to be accessed through circuit differentiation, including implementations based on parallel-in-time perturbations and stochastic estimation techniques, illustrated in \cref{fig:abstract}. Within this framework, time-dependent correlation functions can be reconstructed from a single parametrized circuit family, enabling the extraction of dynamical structure factors and related spectroscopic quantities.\\

The manuscript is organized as follows. In \cref{sec:background}, we provide an overview of the theoretical framework of \gls{lr} theory, laying the foundation for our approach and establishing a unified formalism. In \cref{sec:methods}, we develop a general circuit-differentiation framework for computing \glspl{rgf} and discuss multiple estimator constructions within it, including deterministic and stochastic approaches relevant to near-term devices and extensible to fault-tolerant settings. In \cref{sec:applications}, we review the models and analytical expressions used to illustrate our techniques, and in \cref{sec:results} we present and analyze the corresponding simulation results. Finally, in \cref{sec:conclusions}, we offer concluding remarks. Although we focus on reference toy models as examples, our results remain general and do not rely on any specific physical system. Throughout this work, $\hbar$ is set to $1$.

\section{\label{sec:background}Background}

Let $\uham$ be the Hamiltonian describing a system in isolation and $\tham=\uham+\source\pt{t}$ be the total Hamiltonian for the system weakly coupled to an external time-dependent perturbation (or \emph{source}) $\source\pt{t}$ that can be turned on and off adiabatically. Let us consider the expectation value $\langle o_{\bm{R}}\rangle = \cgs o_{\bm{R}}\gs$ of a local observable $o_{\bm{R}}$, evaluated on the ground state $\gs$ of $\uham$. Here $\bm{R}$ indicates a generic $3$D coordinate in the system and could be, for instance, a specific lattice site if the model is discretized. The source Hamiltonian can be further specified as
\begin{equation}\label{eq:source-definition}
    \source\pt{t^\prime}=\int \mathrm{s}\pt{\bm{r}^\prime, t^\prime} o_{\bm{r}^\prime}\pt{t^\prime}\de\bm{r}^\prime\;,
\end{equation}
assuming that the external perturbation couples linearly to the system operators $o_{\bm{r}^\prime}\pt{t^\prime}$ at positions $\bm{r}^\prime$ through the force function $\mathrm{s}\pt{\bm{r}^\prime, t^\prime}$. Moving into the interaction picture, the operator $o_{\bm{R}}$ evolves under the action of the unperturbed Hamiltonian $\uham$ according to $o^I_{\bm{R}}\pt{t} = \me^{\ramuno\uham t}o_{\bm{R}} \me^{-\ramuno\uham t}$. Meanwhile, the propagator for quantum states is given by the time-ordered exponential
\begin{equation}
    U_\mathrm{s}\pt{t, t_0}=\mathcal{T}\exp\pt{-\ramuno\int^t_{t_0}\source^I\pt{t^\prime}\de t^\prime}\;,
\end{equation}
where the interaction-picture source term is defined as $\source^I\pt{t^\prime} = \me^{\ramuno\uham t^\prime}\source\pt{t^\prime}\me^{-\ramuno\uham t^\prime}$.

\glsunset{fd}
\glsunset{lcp}
\begin{figure*}
    \centering
    \begin{quantikz}
        \lstick[3]{$\gs$} & \qw & \gate[3]{U\pt{\bar{\tau}}} & \gate{R^\beta\pt{\pm\frac{\varepsilon}{2}}} & \gate[3]{U\pt{T-\bar{\tau}}} & \\
        & \qwbundle{\nqubits-2} &&&& \\
        \qw & \qw & \qw & \qw & \qw & \meter{\alpha}
    \end{quantikz}
    \begin{quantikz}
        \lstick[3]{$\gs$} & \qw & \gate[3]{U\pt{\bar{\tau}}} & \gate[style={fill=blue!30}]{R^\beta\pt{\pm\frac{\pi}{2}}} & \gate[3]{U\pt{T-\bar{\tau}}} & \\
        & \qwbundle{\nqubits-2} &&&& \\
        \qw & \qw & \qw & \qw & \qw & \meter{\alpha}
    \end{quantikz}
    \caption{\textbf{Circuit schemes for the finite differences and local circuit perturbation approaches.} Circuit to compute the \gls{rgf} as in \cref{eq:functional-derivative} between the first and last site with a single (local-time) circuit perturbation $R^\beta\pt{\cdot}$. The argument of the rotation gate is a small positive angle $\pm\varepsilon/2$ for the \gls{fd} approach (left) and $\pm\pi/2$ in the \gls{lcp} approach (right). $\gs$ is the initial state, $T$ the total simulation time, $\nperturb$ the number of Trotter steps and $\bar{\tau}=\tau\pt{\bar{n}-1}$, $\tau=T/\nperturb$ (see also main text). $U\pt{\cdot}$ indicates the time evolution operator, $U\pt{0}\equiv\id^{\otimes \nqubits}$ and the quantum register is composed of $\nqubits$ qubits.}
    \label{fig:fd-circuit}
\end{figure*}
\glsreset{lcp}
\glsreset{fd}

The evolution in time of $\langle o_{\bm{R}}\rangle$ induced by the action of the source can then be written as
\begin{equation}\label{eq:obs-evolution}
    \pa{o_{\bm{R}}\pt{t}}_{\mathrm{s}}=\cgs U_\mathrm{s}^\dagger\pt{t, t_0}o^I_{\bm{R}}\pt{t}U_\mathrm{s}\pt{t, t_0}\gs\;.
\end{equation}
Expectation values are from now on taken with respect to the ground state $\gs$, unless otherwise stated. By expanding the propagator to \emph{linear} order in the perturbation $\source$, the change in expectation value of \cref{eq:obs-evolution} up to a final time $T$ is given by the Kubo formula~\cite{kubo_statistical-mechanical_1957}
\begin{equation}\label{eq:first-order-expansion}
    \Delta\pa{o_{\bm{R}}\pt{T}}_{\mathrm{s}}=\ramuno\int^T_{-\infty}\pa{\pq{\source^I\pt{t^\prime}, o^I_{\bm{R}}\pt{T}}}\de t^\prime\;,
\end{equation}
where $\Delta\pa{o_{\bm{R}}\pt{T}}_{\mathrm{s}}=\pa{o_{\bm{R}}\pt{T}}_{\mathrm{s}}-\pa{o_{\bm{R}}}$ denotes the difference between the perturbed and unperturbed expectation values and we introduced the usual commutator $\pq{A,B}=AB-BA$. The integral is extended to $t_0\to-\infty$, assuming that the perturbation is simply turned off in the distant past. \cref{eq:first-order-expansion} fulfills causality, as any change in the expectation value of the observable at time $T$ can only be induced by a perturbation applied at a time $t^\prime < T$. Additionally, if the observable $o_{\bm{R}}$ is normal ordered with respect to the ground-state $\gs$, implying that $\pa{o_{\bm{R}}} = 0$, the first-order fluctuations from \cref{eq:first-order-expansion} simply correspond to the time-evolved expectation value $\Delta\pa{o_{\bm{R}}\pt{T}}_{\mathrm{s}}=\pa{o_{\bm{R}}\pt{T}}_{\mathrm{s}}$. Using \cref{eq:source-definition}, \cref{eq:first-order-expansion} can then be rewritten as
\begin{equation}\label{eq:LRT}
    \pa{o_{\bm{R}}\pt{T}}_{\mathrm{s}}=\int\de\bm{r}^\prime\int^T_{-\infty}\mathcal{G}\pt{\bm{R}, \bm{r}^\prime, T, t^\prime}\mathrm{s}\pt{\bm{r}^\prime, t^\prime}\de t^\prime\;,
\end{equation}
where
\begin{equation}\label{eq:green}
    \mathcal{G}\pt{\bm{R}, \bm{r}^\prime, T, t^\prime}=-\ramuno\Theta\pt{T-t^\prime}\pa{\pq{o^I_{\bm{R}}\pt{T} , o^I_{\bm{r}^\prime}\pt{t^\prime}}}
\end{equation}
is a generalized susceptibility, with the step function $\Theta\pt{T-t^\prime}$ explicitly enforcing causality. This object can be directly identified with the \glspl{rgf} of the observable, an essential quantity that characterizes the system's causal response to external perturbations and provides key information about its dynamical and spectral behavior. Interestingly, \cref{eq:LRT} also shows that the latter can be expressed as the functional derivative
\begin{equation}\label{eq:functional-derivative}
    \mathcal{G}\pt{\bm{R}, \bm{r}^\prime, T, t^\prime} = \frac{\delta \pa{o_{\bm{R}}\pt{T}}_{\mathrm{s}}}{\delta\mathrm{s}\pt{\bm{r}^\prime, t^\prime}} \bigg|_{\mathrm{s}=0} \;.
\end{equation}
As shown in the next section, this formulation provides the basis for the development of several quantum algorithms to compute the \gls{rgf}.

\section{\label{sec:methods}Methods}

The relation in \cref{eq:functional-derivative} enables the definition of a class of estimators for \glspl{rgf} based on circuit derivatives. In this section, we illustrate this framework through concrete examples showing how the required functional derivatives can be accessed using (one or more) parametrized quantum circuits. The working assumption is that a quantum register can be initialized in the ground state $\gs$ and that both the unperturbed and perturbed propagators can be efficiently implemented -- e.g., via a product formula -- in a digital quantum simulation. Moving back to the Schr\"odinger picture, we also assume that -- in the spirit of a Suzuki-Trotter decomposition -- the action of $\uham$ and $\source\pt{t^\prime}$ can be factorized to yield a total propagator up to time $T$ of the form
\begin{equation}\label{eq:general-U}
    U_\mathrm{s}\pt{T, \nperturb}\approx\prod_{n=0}^{\nperturb-1}\pq{\me^{-\ramuno\uham \tau}\me^{-\ramuno\source\pt{n\tau}\tau}}\;,
\end{equation}
where $\tau=T/\nperturb$ and $\source$ is evaluated on a discrete set of time points. We can then construct the functional
\begin{equation}\label{eq:general-F}
\mathcal{F}\pq{\mathrm{s}}\coloneq\pa{U^\dagger_\mathrm{s}\pt{T,\nperturb}o_{\bm{R}} U_\mathrm{s}\pt{T,\nperturb}}\;,
\end{equation}
which corresponds to the digital quantum simulation of the expectation value $\pa{o_{\bm{R}}\pt{T}}$ with a specific choice of the force function $\mathrm{s}\pt{\bm{r}^\prime, t^\prime}$ across positions $\bm{r}^\prime$ and times $t^\prime \in\pq{0,T}$ for a total evolution time $T$.
\begin{figure*}
    \centering
    \begin{quantikz}
      \lstick[3]{$\gs$} & \gate[style={fill=green!30}]{R^\beta\pt{\varepsilon\eta^{\pt{0}}_p}} & \gate[3]{U\pt{\tau}} & \gate[style={fill=green!30}]{R^\beta\pt{\varepsilon\eta^{\pt{2}}_p}} & \gate[3]{U\pt{\tau}} & \ \ldots\ & \gate[style={fill=green!30}]{R^\beta\pt{\varepsilon\eta^{\pt{\nperturb-1}}_p}} & \gate[3]{U\pt{\tau}} & \\
      & \qwbundle{\nqubits-2} &&&& \ \ldots\ &&& \\
      \qw & \qw & \qw & \qw & \qw & \ \ldots\ & \qw & \qw & \meter{\alpha}
    \end{quantikz}
    \caption{\textbf{Circuit scheme for the simultaneous circuit perturbation approach.} Circuit to compute the \gls{rgf} as in \cref{eq:functional-derivative} between the first and last site with $\nperturb$ circuit perturbations $R^\beta\pt{\varepsilon\eta^{\pt{k}}_p}$. Each element $\eta^{\pt{k}}_p$ of the $\nperturb$-dimensional random vector $\bm{\eta}_p$ is sampled from a Rademacher distribution. For the notation, see also the caption to \cref{fig:fd-circuit}}.
    \label{fig:spsa-circuit}
\end{figure*}
\subsection{Local circuit perturbation}\label{sec:lcp}

By using a Dirac delta in both position and time as a test function, the functional derivative of $\mathcal{F}$ with respect to $\mathrm{s}$ can formally be expressed as
\begin{equation}\label{eq:finite-diff-functional}
\frac{\delta \mathcal{F}\pq{\mathrm{s}}}{\delta\mathrm{s}}=\lim_{\varepsilon\to 0}\frac{\mathcal{F}\pq{\mathrm{s}+\varepsilon\delta\pt{\bm{r}^{\prime}-\bm{r}}\delta\pt{t^\prime - t}}-\mathcal{F}\pq{\mathrm{s}}}{\varepsilon}\;.
\end{equation}
Recalling from \cref{eq:functional-derivative} that $\mathcal{G}$ is given by the above expression evaluated around a constant null perturbation $\mathrm{s}=0$, we can approximate it via the \gls{fd} expression
\begin{align}\label{eq:finite-diff-chi}
    \mathcal{G}\pt{\bm{R}, \bm{r}, T, t} & \approx \Big[\pa{U^\dagger_{\delta\mathrm{s}}\pt{T,\nperturb}o_{\bm{R}} U_{\delta\mathrm{s}}\pt{T,\nperturb}}\nonumber \\
    &-\pa{U_0^\dagger\pt{T,\nperturb}o_{\bm{R}} U_0\pt{T,\nperturb}}\Big]\varepsilon^{-1}\;,
\end{align}
where $\delta\mathrm{s} = \varepsilon\delta\pt{\bm{r}^{\prime}-\bm{r}}\delta\pt{t^\prime - t}$; with some abuse of notation, we identify from here on $U_{\delta\mathrm{s}}\equiv U_{\varepsilon}$. The time evolution in the first line of \cref{eq:finite-diff-chi} contains a contribution generated by the space- and time-local perturbation
\begin{equation}
H_\varepsilon\pt{t^\prime; \bm{r}, t}=\varepsilon\delta\pt{t^\prime-t}o_{\bm{r}}\;,
\end{equation}
with $\varepsilon \ll 1$, i.e., a small kick applied to the system at position $\bm{r}$ and time $t$, whose parametric dependency is indicated after the semicolon. The second term of \cref{eq:finite-diff-chi} is instead the ground-state expectation value $\pa{o_{\bm{R}}}$, which remains constant under the unperturbed evolution assuming negligible Suzuki-Trotter discretization errors. Notice that, while in \cref{eq:finite-diff-functional} we used the so-called forward definition of the \gls{fd} ratio, the accuracy can be improved from $\mathcal{O}\pt{\varepsilon}$ to $\mathcal{O}\pt{\varepsilon^2}$ by using a symmetric increment of $\pm\varepsilon/2$. Such higher-order version (see \cref{fig:fd-circuit}) will be adopted henceforth.

To be more concrete, assume that $o_{\bm{R}}=\sigma_R^\alpha$ and $o_{\bm{r}}=\sigma^\beta_{r}$ are, for instance, single qubit Pauli operators acting on a one-dimensional spin-$1/2$ lattice. In a quantum circuit representation, the first term on the RHS of \cref{eq:finite-diff-chi} is then obtained by introducing a single-qubit rotation gate around the axis $\beta$ on the $r$-th qubit at the $\bar{n}$-th time step (with $t=\bar{n}\tau$) and measuring $\sigma^\alpha_R$ at the final time $T$, see \cref{fig:fd-circuit} (left). The second term is, as already mentioned, simply the unperturbed ground state expectation value of $\sigma^\alpha_R$. For $t=0$, this protocol essentially corresponds to the algorithm proposed in Ref.~\cite{kokcu_linear_2024} in which, similarly to an actual \gls{lr} experiment, a system is initialized in the ground state and locally driven with a small external field.

In practice, the \gls{fd} method applied in \cref{eq:finite-diff-chi} is highly susceptible to numerical instabilities, particularly in the context of quantum computation: in realistic scenarios, the calculation of expectation values is affected by both hardware and statistical noise. While the former can typically be addressed through error suppression and mitigation techniques~\cite{cai_quantum_2023,berg_probabilistic_2023,kim_scalable_2023,miessen_benchmarking_2024}, the latter often requires substantial sampling efforts to achieve sufficiently high precision for a reliable application of this approach.

Interestingly, a more robust solution is available in many concrete use cases via a targeted choice of the perturbation parameter. In fact, one may notice that the expression for $\mathcal{G}$ in \cref{eq:finite-diff-chi} is simply an approximation of the partial derivative $\partial \fdfunc/\partial\varepsilon$, evaluated at $\varepsilon=0$, of the function
\begin{equation}\label{eq:LCP-U}
    \fdfunc\pt{\varepsilon}=\pa{U_\varepsilon^\dagger\pt{T,\nperturb}o_{\bm{R}} U_\varepsilon\pt{T,\nperturb}}\;,
\end{equation}
with 
\begin{equation}
    U_\varepsilon\pt{T,\nperturb}=\me^{-\ramuno\uham\pq{T-\bar{\tau}}}\me^{-\ramuno\varepsilon o_{\bm{r}}\pt{\bar{n}\tau}}\me^{-\ramuno\uham\bar{\tau}}
\end{equation}
and $\bar{\tau}=\tau\pt{\bar{n}-1}$. The quantity $\fdfunc\pt{\varepsilon}$ can be directly interpreted as the output produced by an $\varepsilon$-parametrized quantum circuit~\cite{mangini2021quantum}. If $H_\varepsilon$ is sufficiently simple, one can then apply what is known as the \gls{psr}~\cite{mitarai_quantum_2018,schuld2019evaluating} to obtain an \emph{exact} expression for $\partial\fdfunc/\partial\varepsilon$ in terms of two evaluations of $\fdfunc$ on macroscopically different choices of the input. In particular, when $U_\varepsilon\pt{\tau}=\me^{-\ramuno\varepsilon o_{\bm{r}}\pt{\bar{n}\tau}}$ is generated by a single-qubit Pauli operator, the finite shift values are simply $\pm\pi/2$ and \cref{eq:finite-diff-chi} is replaced by
\begin{equation}\label{eq:psr-expression}
    \mathcal{G}\pt{\bm{R}, \bm{r}, T, t}=1/2\pq{\fdfunc\pt{\pi/2}-\fdfunc\pt{-\pi/2}}\;.
\end{equation}
This approach can be formalized by demonstrating that the derivation of the \gls{psr} \emph{naturally} results in an analytical expression identical to the \gls{rgf}, as we show in \cref{appendix:parameter-shift}. It is interesting to note that the two contributions in \cref{eq:psr-expression} are equal and opposite due to the symmetry in the product of the expectation values of the \glspl{rgf}~\cite{baez_dynamical_2020}. Consequently, flipping the sign of the angle results in a sign flip of the \glspl{rgf}. A schematic description of this technique, which we will refer to as \gls{lcp} in the following, is depicted in \cref{fig:fd-circuit} (right). The \gls{lcp} result, which is consistent with the derivation based on Ramsey interferometry reported in Ref.~\cite{baez_dynamical_2020}, can in principle be generalized to more complex $H_\varepsilon$ generators using linear combination of unitaries~\cite{schuld2019evaluating}, albeit at the cost of (locally) reintroducing one ancilla. In this case the latter, unlike in the Hadamard test, only needs to be connected to the qubit(s) associated with the first perturbation and not with the final readout one(s). Since the initial state is the ground state of $\uham$, any temporal evolution before the perturbation at $t$ will be trivial as long as Suzuki-Trotter discretization errors can be neglected. In practice, this allows us to remove the block $U\pt{\bar{\tau}}$ in \cref{fig:fd-circuit} (right), which in turn leads to shorter circuits for short times when applying \gls{lcp}.

\subsection{Simultaneous circuit perturbation}\label{sec:SCP}

The \gls{lcp} procedure outlined above is particularly convenient, compared to standard ancilla-based quantum algorithms used to evaluate dynamical correlations, since it does not require a significant increase in circuit complexity or additional qubit-qubit connectivity (as long as the target observables are sufficiently local) with respect to the basic digital quantum simulation of the target model. Moreover, if different sets of qubits encode for different positions $\bm{R}$ (as it is the case, e.g., for spin lattice simulations), all local $o_{\bm{R}}$ operators commute and can therefore be measured in parallel, thus requiring only a single circuit instance to reconstruct $\mathcal{G}\pt{\bm{R}, \bm{r}, T, t}$ between all possible choices of $\bm{r}$ and a fixed position $\bm{r}$. However, $t$ is fixed as the point in time at which the perturbation gate is inserted, such that multiple circuits are required to track the dynamics of the correlations. By relaxing the requirement for an analytic gradient, it is possible to devise a new algorithm that satisfies these conditions.

In particular, for fixed $\bm{R}$, $\bm{r}$ and $T$, the vector
\begin{equation}
    \bm{\mathcal{G}} = \begin{pmatrix}
        \mathcal{G}\pt{\bm{R}, \bm{r}, T, t_0} \\ \mathcal{G}\pt{\bm{R}, \bm{r}, T, t_1} \\ \vdots \\ \mathcal{G}\pt{\bm{R}, \bm{r}, T, t_{\nperturb-1}}
    \end{pmatrix} = \left.\begin{pmatrix}
         \delta \mathcal{F}\pq{\mathrm{s}}/\delta\mathrm{s}\pt{\bm{r}, t_0} \\ \delta \mathcal{F}\pq{\mathrm{s}}/\delta\mathrm{s}\pt{\bm{r}, t_1} \\ \vdots \\ \delta \mathcal{F}\pq{\mathrm{s}}/\delta\mathrm{s}\pt{\bm{r}, t_{\nperturb-1}}
    \end{pmatrix}\right|_{\mathrm{s}=0}\;,
\end{equation}
for $t_n=n\tau$, is equivalent to the quantum circuit gradient 
\begin{equation}
    \bm{\mathcal{G}}\equiv \left.\nabla \fdfunc\pt{\bm{\varepsilon}}\right|_{\varepsilon=0}\;.
\end{equation}
Here, we have introduced the natural generalization of the $\fdfunc\pt{\varepsilon}$ function from \cref{eq:LCP-U} to an $\nperturb$-parameter quantum circuit
\begin{equation}\label{eq:SCP-F}
    \fdfunc\pt{\bm{\varepsilon}} = \pa{U_{\bm{\varepsilon}}^\dagger\pt{T,\nperturb}o_{\bm{R}} U_{\bm{\varepsilon}}\pt{T,\nperturb}}\;,
\end{equation}
where $\bm{\varepsilon}=\pt{\varepsilon_0,\ldots,\varepsilon_{\nperturb-1}}$ and the corresponding perturbation function is
\begin{equation}
    \bm{\delta\mathrm{s}} = \begin{pmatrix} \varepsilon_0 \delta\pt{\bm{r}^{\prime}-\bm{r}}\delta\pt{t^\prime - t_0}\\ 
    \vdots\\
    \varepsilon_{p-1} \delta\pt{\bm{r}^{\prime}-\bm{r}}\delta\pt{t^\prime - t_{p-1}} \end{pmatrix}
\end{equation}
with
\begin{equation}\label{eq:SCP-U}
    U_{\bm{\varepsilon}}\pt{T,\nperturb}=\prod_{n=0}^{\nperturb-1}\me^{-\ramuno\uham\tau}\me^{-\ramuno\varepsilon_n o_{\bm{r}}\pt{n\tau}}\;.
\end{equation}

Inspired by optimization methods that rely on gradient estimators, we adopt the stochastic approach underlying \gls{spsa}~\cite{spall_multivariate_1992} to obtain, from a single circuit template and multiple randomized perturbations acting in parallel, an estimate of the whole gradient $\nabla\fdfunc\pt{\bm{\varepsilon}}$. Instead of local in time perturbations, the \gls{spsa} estimator is constructed by sampling realizations of a $\nperturb$-dimensional vector $\bm{\eta}\in\pg{\pm 1}^{\nperturb}$ uniformly at random, setting $\bm{\varepsilon} = \varepsilon\bm{\eta}$ and measuring the expectation value of $o_{\bm{R}}$ from the resulting circuit. In general, the latter can be constructed by taking $S$ \emph{shots}, meaning that the circuit is executed $S$ times for each fixed choice of $\bm{\eta}$. If we denote with $v^{\pt{k}}_p$ the $k$-th element of the $p$-th random realization of a vector $\bm{v}$, the $k$-th component of the \gls{spsa} gradient estimator constructed by sampling $P$ perturbation directions (i.e. realizations of the $\bm{\eta}$ vector) and using for each of those $S$ circuit calls is given in its general form by
\begin{equation}\label{eq:gradient-SPSA}
    \grad=\frac{1}{P}\sum^{P-1}_{p=0}\frac{1}{S}\sum^{S-1}_{s=0}\frac{\exfunc_s\pt{\theta+\varepsilon\bm{\eta}_p}-\exfunc_s\pt{\theta-\varepsilon\bm{\eta}_p}}{2\varepsilon}\eta^{\pt{k}}_p\;,
\end{equation}
where $\exfunc_s$ is the estimator of a target observable (or cost function value in an optimization setting) obtained from the $s$-th shot and with the parameter vector $\theta$ perturbed by $\pm\varepsilon\bm{\eta}_p$ along the random direction $\bm{\eta}_p$. We note that $P$ is independent of the total evolution time $T$ and number of Trotter steps $N$. In addition, the number of entries of the vector $\bm{\eta}$ may be, in general, $|\bm{\eta}|\leq N$, where the equality is obtained if one allows (as we do in this work) for one perturbation per Trotter step. While $\varepsilon$ is typically updated iteratively in optimization problems, in our case it remains fixed at a finite value $<1$. Here, $\exfunc_s$ is an estimate of $o_{\bm{R}}$, $\theta=0$ and the total shot budget is $\mathcal{S}=PS$. We refer to this parallel in time algorithm as the \gls{scp} technique. For the simple case of single-spin perturbations, the structure of the quantum circuit is shown in \cref{fig:spsa-circuit}, where unperturbed Trotter time evolution steps are interleaved with single-qubit rotations of random angles $\varepsilon^{\pt{k}}_p=\varepsilon\eta^{\pt{k}}_p$.

It is now easy to see that, for a compatible set of operators $o_{\bm{R}}$, we can simultaneously estimate $\mathcal{G}\pt{\bm{R}, \bm{r}, T, t}$ between a fixed pair $\pt{\bm{r},T}$ and multiple choices of $\bm{R}$ and $t$, all with a single circuit template and a single sampling experiment where only the perturbation vectors $\pg{\bm{\eta}_p}^{P-1}_{p=0}$ need to be randomized. Given that the Hamiltonian $\uham$ is time-independent, the \gls{rgf} depends only on the time difference between the perturbation and the response, rather than on their absolute times. The time-translational invariance allows the \gls{scp} approach to reconstruct the entire $\mathcal{G}\pt{\bm{R}, \bm{r}, T, T - t}$ trace using a single circuit. In contrast, the \gls{lcp} method requires a separate experiment for each point on the curve, albeit with typically shorter evolution times. As a result, \gls{scp} is particularly advantageous when dense or long-time sampling of \glspl{rgf} is required, whereas deterministic approaches such as \gls{lcp} may be preferable when only a small number of short-time values are needed due to their reduced circuit depth.

From a practical perspective, stochastic estimators such as \gls{scp} naturally align with the capabilities of contemporary quantum hardware, where high sampling rates and parametric circuit compilation techniques are increasingly available, particularly on superconducting platforms~\cite{fischer2024dynamical}. This is especially relevant in light of the fact that, as shown in \cref{appendix:variance}, the optimal estimator variance is obtained by using $S=1$, i.e. a single shot per randomization of the perturbation vector, with sampling error following the usual Monte Carlo inverse square-root scaling.

More generally, this applies to all approaches within the broader class of fixed-template circuit-differentiation, i.e. methods characterized by constant circuit depth and a consistent layer-wise structure across time. Besides \gls{scp}, this also includes specific implementations of \gls{lcp} based on a single maximum-time, fixed-depth circuit template of the same form used in the \gls{scp} construction, where only one perturbation parameter is set to a non-zero value at a time. This class of circuits also exhibits favorable properties for noise-learning-based error mitigation strategies, including probabilistic error cancellation~\cite{berg_probabilistic_2023} and global calibration~\cite{farrell_scalable_2024,urbanek_mitigating_2021}. Since identical circuit instances are used for all time points, noise properties are shared uniformly across the reconstructed Green's function, leading to more homogeneous variances and facilitating mitigation methods based on operator renormalization or affine transformations. These features are particularly relevant for dynamical correlation functions, where subsequent analysis often relies on global operations applied to the full time series, such as Fourier transforms.

\subsection{Extensions of the circuit-differentiation framework}\label{sec:extensions}
Beyond the specific estimators discussed above, the circuit-differentiation formulation introduced here naturally admits further generalizations. In particular, once \glspl{rgf} are expressed as derivatives of a scalar circuit output, a broad class of gradient-estimation techniques developed for parametrized quantum circuits becomes directly applicable. As an illustrative example, we consider the gradient-estimation method of Ref.~\cite{huggins_nearly_2022}, which accesses circuit gradients by encoding an objective function $f\pt{\mathrm{s}}$ into a probability oracle. In \cref{appendix:extension-gradient}, we define such a function via \cref{eq:general-F} and we show that it satisfies the analyticity and boundedness conditions required by this framework, enabling a natural connection to coherent gradient-estimation methods in the fault-tolerant regime. From this perspective, the advantage observed for \gls{scp} (see \cref{sec:results}) may be viewed as a near-term, heuristic analogue of the same parallel-estimation principle, which in the fault-tolerant setting can be formalized to yield a provable quadratic improvement -- reducing the cost of estimating $M$ response coefficients up to precision $\varepsilon$ from $\mathcal{O}\pt{M/\varepsilon}$ to $\mathcal{O}\pt{\sqrt{M}/\varepsilon}$, up to logarithmic factors.

In our setting, due to the commutator structure of the \glspl{rgf}, $f\pt{\mathrm{s}}$ is an expectation value rather than the imaginary part of an overlap between quantum states as considered in Eq.~(8) of Ref.~\cite{huggins_nearly_2022}. Since Pauli expectations can be mapped directly to measurement probabilities, the task of constructing a probability oracle becomes straightforward. Specifically, we seek a circuit that produces an ancilla qubit whose measurement outcome occurs with probability $f\pt{\mathrm{s}}$. Such an oracle is given by
\begin{equation}\label{eq:probability-oracle}
    A\pt{\mathrm{s}}\coloneq\Pi_P\pt{\mathbb{1}\otimes R_P}\pt{\mathbb{1}\otimes U_\mathrm{s}}\pt{\mathbb{1}\otimes U_{\psi_0}}\;,
\end{equation}
where $U_{\psi_0}\ket{0}^{\otimes\nqubits}=\gs$ (with $\nqubits$ denoting the total number of register qubits) and $U_\mathrm{s}$ is the perturbed evolution defined in \cref{eq:general-U}. For concreteness, we consider $o_{\bm{R}}$ to be a Pauli string $P$ with support $S\pt{P}\subseteq\pg{0,\dots,\nqubits-1}$ and weight $w=|S\pt{P}|$. The unitary $R_P$ implements a basis-change that diagonalizes $o_{\bm{R}}$, such that $R_PPR^\dagger_P=Z^{\otimes w}\otimes\mathbb{1}^{\nqubits-w}$. The parity extraction unitary $\Pi_P$ coherently computes the parity of the corresponding $Z$-measurement outcomes into an ancilla qubit $a$,
\begin{equation}
\Pi_P\coloneq\prod_{j\in S\pt{P}}\text{CNOT}_{j\to a}\;.
\end{equation}
Starting from the ancilla state $\ket{0}_a$, applying $\Pi_P$ will store the parity of the involved $Z$-measurement bits in the ancilla. In fact, given an input $\ket{\Psi}=\ket{0}_a\ket{0^\nqubits}$, \cref{eq:probability-oracle} yields
\begin{equation}
    A\pt{\mathrm{s}}\ket{\Psi}=\sqrt{f\pt{\mathrm{s}}}\ket{0}_a\ket{\phi_0\pt{\mathrm{s}}}+\sqrt{1-f\pt{\mathrm{s}}}\ket{1}_a\ket{\phi_1\pt{\mathrm{s}}}
\end{equation}
for some normalized states $\ket{\phi_0\pt{\mathrm{s}}}$ and $\ket{\phi_1\pt{\mathrm{s}}}$. In particular, the probability of obtaining the outcome $a=0$ upon measurement of the ancilla is $\mathrm{Pr}\pt{a=0}= f\pt{\mathrm{s}}$. The ancilla construction defined by \cref{eq:probability-oracle} relies on a coherent readout of a Pauli measurement instead of a Hadamard-test on the full time evolution unitary. In this sense, the definition in \cref{eq:general-F} identifies a class of parameterized circuits that avoids a substantial portion of the controlled operations required in Ref.~\cite{huggins_nearly_2022} for correlation-function estimation, while still requiring coherent access to $A\pt{\mathrm{s}}$ and its inverse.

\section{\label{sec:applications}Applications}

An important application of our proposed quantum algorithms is the efficient computation of the \gls{dsf}, a key quantity for investigating the dynamical properties of condensed matter systems. As mentioned earlier, all target models are assumed to exhibit time-translational invariance, i.e., their Hamiltonians are time-independent. In addition, the Hamiltonian $\uham$ is taken to be spatially translationally invariant, implying that coupling constants and interactions are identical across all sites.

The \gls{dsf} provides insight into the low-energy excitation spectrum and ground-state correlations of a medium. For a system of size $\nqubits$, given $\int_{\mathcal{R},t} \equiv \sum_{\bm{R},\bm{r}} \int_{-\infty}^{\infty}\de t$, it is defined by
\begin{equation}
     \mathcal{S}\pt{\bm{q},\omega}=\frac{1}{2\pi \nqubits}\int_{\mathcal{R},t}\me^{-\ramuno\omega\pt{T-t}}\me^{-\ramuno\bm{q}\cdot\pt{\bm{R}-\bm{r}}}\mathcal{C}\pt{\bm{R}, \bm{r}, T, t}\label{eq:dsf}\;,
\end{equation}
i.e., as the Fourier transform in both space and time of dynamical correlation functions of the form
\begin{equation}\label{eq:dcs}
    \mathcal{C}\pt{\bm{R}, \bm{r}, T, t}=\pa{o_{\bm{R}}\pt{T}o_{\bm{r}}\pt{t}}\;.
\end{equation}
The \gls{dsf} can be easily measured, for instance, in inelastic neutron scattering experiments, and thus represents a direct link between theoretical models and experimental observations. However, computing the \gls{dsf} for many body quantum systems using classical simulation techniques is, in general, a very demanding task. In fact, it has been shown that even approximate estimates of the \gls{dsf} for arbitrary local Hamiltonians are BQP-hard to construct~\cite{baez_dynamical_2020}, thus making the \gls{dsf} an ideal target of both digital and analog quantum simulation protocols~\cite{chiesa_quantum_2019,tacchino2020quantum,bauer2025progress}. Whenever the system is in thermal equilibrium at inverse temperature $\beta$ and \gls{lr} theory applies, the \gls{dsf} can be expressed in terms of the imaginary part of the \gls{rgf} via the fluctuation-dissipation theorem~\cite{RKubo_1966,kubo2012statistical}, namely
\begin{equation}
    \mathcal{S}\pt{\bm{q},\omega} =\pt{2n\pt{\omega}\pm 1}\imag\pq{\mathcal{G}\pt{\bm{q}, \omega}}\;.
\label{eq:fluctuation-dissipation}
\end{equation}
Here, $n\pt{\omega}=\pt{\me^{\beta\omega}\mp 1}^{-1}$ is the Bose-Einstein (Fermi-Dirac) distribution function and $\mathcal{G}\pt{\bm{q}, \omega}$ is the Fourier transform (in time and space) of $\mathcal{G}\pt{\bm{r}, \bm{r}^\prime, t, t^\prime}$. This relation formalizes the intuitive idea that if a system's susceptibility to external changes is large, its fluctuations around equilibrium will be equally large. In fact, the imaginary part of the response function constitutes a measure of the rate of energy dissipation in the system.

\subsection{Heisenberg spin-\texorpdfstring{$1/2$ model}{}}\label{sec:Heis}

As a first target, we consider the one-dimensional Heisenberg model, consisting of a chain of $L$ spin-$1/2$ particles coupled via nearest-neighbor exchange interactions
\begin{equation}\label{eq:heisenberg}
    \tham=\underset{\alpha=x,y,z}{\sum^{\nqubits-1}_{r=0}} J^\alpha_r\sigma^\alpha_r\sigma^\alpha_{r+1}\;,
\end{equation}
where the $\sigma^\alpha$ are Pauli operators, $J^\alpha_r$ is the coupling strength between lattice sites $r$ and $r+1$ along the spin axis $\alpha$ (set for our simulations uniformly to $J=1$) and $\sigma^\alpha_{\nqubits}=\sigma^\alpha_0$. Furthermore, to approximate the unitary dynamics we employ a first-order Trotter decomposition, which factorizes the time evolution operator into a sequence of two-qubit gates of the form
\begin{equation}
    \me^{-\ramuno\tham\tau} = \prod_{\alpha = x, y, z} \prod_{r=0}^{\nqubits-1} \me^{-\ramuno J_r^\alpha \sigma_r^\alpha \sigma_{r+1}^\alpha \tau} + \mathcal{O}\pt{\tau^2}\;.
\end{equation}
For the Heisenberg model, the definition of the \gls{rgf} specializes to
\begin{equation}\label{eq:green-spins}
    \mathcal{G}^{\alpha\beta}\pt{R,T}=-\frac{\ramuno}{2}\Theta\pt{T}\pa{\pq{{\sigma^\alpha_R\pt{T} ,\sigma^\beta_0\pt{0}}}}\;,
\end{equation}
where we have set $o_r\equiv\sigma^\alpha_r$ and, leveraging both spatial and temporal translational invariance, we have chosen the site $r=0$ and time $t=0$ as a reference. It is also easy to show that, when using spin operators (which are Hermitian), the \gls{rgf} is equal to the imaginary part of the corresponding dynamical correlation function
\begin{equation}\label{eq:green-is-imag}
    \mathcal{G}^{\alpha\beta}\pt{R, T}=\imag\pq{\mathcal{C}^{\alpha\beta}\pt{R, T}} = \imag\pq{\pa{{\sigma^\alpha_R\pt{T}\sigma^\beta_0\pt{0}}}}\;.
\end{equation}
By expanding $\mathcal{C}^{\alpha\beta}\pt{r, t}$ on a basis of Hamiltonian eigenvectors $|e\rangle$ with corresponding eigenfrequencies $\omega_e$, we get
\begin{align}
    \mathcal{C}^{\alpha\beta}\pt{r, t}&=\sum_e \braket{\psi_0|\sigma^\alpha_r|e}\braket{e|\sigma^\beta_{0}|\psi_0}\me^{-\ramuno \omega_e t}\nonumber \\
    &= \sum_e\pt{a_{r,e}^{\alpha\beta} + \ramuno b_{r,e}^{\alpha\beta}}\me^{-\ramuno \omega_e t}\nonumber \\
    &= \sum_e\pt{a_{r,e}^{\alpha\beta}\cos\omega_e t + b_{r,e}^{\alpha\beta}\sin\omega_e t}\nonumber \\
    &+\ramuno \sum_e\pt{b_{r,e}^{\alpha\beta}\cos\omega_e t-a_{r,e}^{\alpha\beta}\sin\omega_e t}\;,
\end{align}
where $a_{r,e}^{\alpha\beta}$ ($b_{r,e}^{\alpha\beta}$) are the real (imaginary) parts of the overlap coefficient $\langle\psi_0|\sigma^\alpha_r|e\rangle\langle e|\sigma^\beta_{0}|\psi_0\rangle$. Since the eigenvectors of the Heisenberg model are real, we can a priori set $b_{r,e}^{\alpha\beta} = 0$ for any combination of indices~\cite{chiesa_quantum_2019}. We therefore conclude, using \cref{eq:green-is-imag}, that the desired \gls{rgf} consists only of sinusoidal (i.e., odd) terms,
\begin{equation}\label{eq:fitting-function}
    \mathcal{G}^{\alpha\beta}\pt{r, t}=\textstyle\sum_ea_{r,e}^{\alpha\beta}\sin\omega_e t\;,
\end{equation}
whose Fourier transform in time is then purely imaginary
\begin{equation}
    \mathcal{F}_t\pt{\mathcal{G}^{\alpha\beta}\pt{r, t}}\propto\ramuno\pi\sum_ea_{r,e}^{\alpha\beta}\delta\pt{\omega-\omega_e}\;,
\end{equation}
where we have selected the term with physical (positive) energies $\omega_e\ge0$. Finally, taking the Fourier transform in space introduces an oscillatory factor, and by applying \cref{eq:fluctuation-dissipation} and extracting the imaginary part, we find that only the even terms in $r$ survive, so that the \gls{dsf} can be expressed as
\begin{equation}\label{eq:dsf-final-expression}
    \mathcal{S}^{\alpha\beta}\pt{q,\omega}=\textstyle\sum_{r,e}a_{r,e}^{\alpha\beta}\cos\pt{qr}\delta\pt{\omega-\omega_e}
\end{equation}
up to Fourier normalization factors. \cref{eq:dsf-final-expression} is reminiscent of the Lehmann representation of the \gls{dsf} and constitutes the expression we will use to practically compute $\mathcal{S}^{\alpha\beta}\pt{q,\omega}$ for the numerical simulations in \cref{sec:results}.
\begin{figure*}
\includegraphics{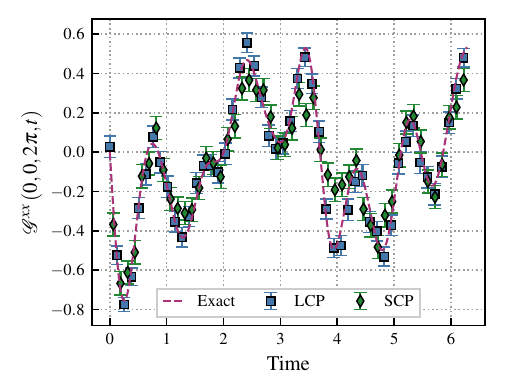}
\includegraphics{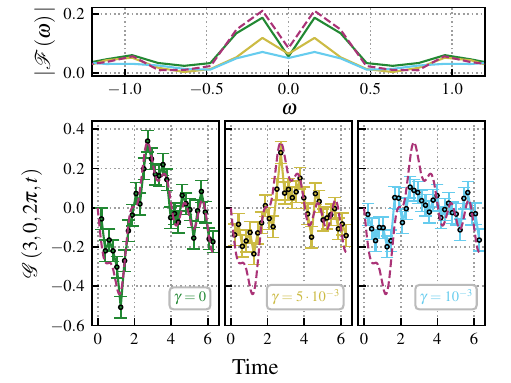}
\caption{\textbf{Comparison of the dynamics generated by local and simultaneous circuit perturbation.} For both plots, the expected dynamics is shown as the dashed magenta line. \textit{Left}. Bosonic Green's function as in \cref{eq:green} for a $10$-spin Heisenberg ring, $\alpha=\beta=x$. The expected dynamics with $\nperturb=100$ Trotter steps is compared against the results produced with both the circuits in \cref{fig:fd-circuit} (\gls{lcp}) and \cref{fig:spsa-circuit} (\gls{scp}). The same shot budget is evenly distributed across each point of the \gls{lcp} curve, whereas the entire budget is allocated for \gls{scp}. \textit{Right}. Fermionic Green's function as in \cref{eq:fgreen} for a $4$-site $1$D periodic Hubbard model, $J=1$ and $U=5$. The expected dynamics with $\nperturb=30$ Trotter steps is compared with the \gls{scp} method at different values of the depolarizing parameter $\gamma$. The top plot shows the Fourier transform of the dynamics (amplitude vs. frequency), with decreasing peaks for larger noise values. For both simulations, $T=2\pi$ and error bars indicate one standard deviation. For details on their calculation, see also \cref{fig:variance}.} \label{fig:together}
\end{figure*}

\subsection{Fermi-Hubbard model}

We now expand our analysis to include fermionic systems. As a test bench, we employ the Fermi-Hubbard model independently introduced in~\cite{hubbard_electron_1963,kanamori_electron_1963,gutzwiller_effect_1963}, describing interacting fermions on a lattice via the Hamiltonian
\begin{equation}\label{eq:hubbard}
    \tham = -J \sum_{\sigma, \braket{\mu\nu}}
        \pt{a^\dagger_{\sigma, \mu} a_{\sigma, \nu} + a^\dagger_{\sigma, \nu} a_{\sigma, \mu}}+ U \sum_\mu n_{\alpha, \mu} n_{\beta, \mu}\;,
\end{equation}
where $J$ and $U$ are respectively the competing kinetic energy (hopping) and on-site Coulomb repulsion coefficients and the sums run over both spin species $\alpha$, $\beta$ and nearest-neighbor sites $\mu$, $\nu$.

For fermionic systems, the formalism introduced in \cref{sec:methods} becomes necessarily more involved. Following the same idea as in the \gls{lr} setting, one may add to the Hamiltonian a perturbation
\begin{equation}\label{eq:fermionic-source}
    \tham_{\mathrm{s}}=\int\de \bm{r}^\prime\pt{\mathrm{s}_1\pt{\bm{r}^\prime, t^\prime}a_{\bm{r}^\prime} + \mathrm{s}_2\pt{\bm{r}^\prime, t^\prime} a^\dagger_{\bm{r}^\prime}}\;,
\end{equation}
where the usual creation and annihilation operators at position $\bm{r}^\prime$, $a^\dagger_{\bm{r}^\prime}$ and $a_{\bm{r}^\prime}$ are introduced. In second quantization, this term can either inject a particle at position $\bm{r}^\prime$ or remove one from that same position. In order for \cref{eq:fermionic-source} to be part of a Hamiltonian, it must have a well-defined classical limit, $\hbar\to0$, where the local Hamiltonian terms effectively commute at large distances and do not introduce non-commuting behavior over long ranges~\cite{giamarchi2013manybody}. However, if $\mathrm{s}_1$ and $\mathrm{s}_2$ are ordinary commuting fields, the creation and annihilation operators fail to commute, regardless of the distance. This issue can be resolved by promoting $\mathrm{s}_1$ and $\mathrm{s}_2$ to Grassmann variables, which anti-commute. This translates into an anti-commutator (rather than a commutator) in the fermionic \gls{rgf} expression,
\begin{equation}\label{eq:fgreen}
    \mathcal{G}\pt{\bm{R}, \bm{r}, T, t}=-\ramuno\Theta\pt{T-t}\pa{\pg{a_{\bm{R}}\pt{T}, a^\dagger_{\bm{r}}\pt{t}}}\;,
\end{equation}
with $\pg{A,B}=AB+BA$. However, the protocol described in \cref{sec:methods} is only capable of computing commutators and thus requires adaptation. A possible approach is suggested by the auxiliary operator method used in Ref.~\cite{kokcu_linear_2024} which makes the problem of calculating the anti-commutator in \cref{eq:fgreen} equivalent to that of calculating the commutator
\begin{equation}\label{eq:auxiliary-operator}
    \mathcal{G}\pt{\bm{R}, \bm{r}, T, t}=\frac{\ramuno}{\lambda}\Theta\pt{T-t}\pa{\pq{\mathcal{P}a_{\bm{R}}\pt{T}, a^\dagger_{\bm{r}}\pt{t}}}\;,
\end{equation}
where $\mathcal{P}$ is a time-independent operator for which it holds $\mathcal{P}\gs=\lambda\gs$ and $\pg{a^\dagger\pt{t}, \mathcal{P}}=0\ \forall t$. As noted in Ref.~\cite{kokcu_linear_2024}, these assumptions are satisfied, for parity-conserving Hamiltonians, by the parity operator.

The final step involves the mapping of the operators to qubits, for which we make use of the \gls{jw} transformation~\cite{jordan_ber_1928}, that for creation and annihilation operators yields
\begin{align}\label{eq:cre-ann-jw}
    a_r^\dagger&=\mathcal{Z}_r\frac{\sigma_r^x+\ramuno \sigma_r^y}{2} &
    a_r&=\mathcal{Z}_r\frac{\sigma^x_r-\ramuno\sigma^y_r}{2}\;,
\end{align}
where we have denoted with $\mathcal{Z}_r=\prod_{j<r}\sigma^z_j$ the attached \gls{jw}-strings. Note that both operators in \cref{eq:cre-ann-jw} anti-commute with the parity operator $\mathcal{P}=Z_0\dots Z_{2\nqubits-1}$, where the factor of $2$ takes care of the possibility of having different spin species. In practice then, two quantum circuits with distinct $\mathcal{Z}_r\sigma^{x/y}_r$ perturbations are therefore required for the measurement of the \gls{rgf}. With these adjustments, both the \gls{lcp} and \gls{scp} protocols can be employed in the fermionic case, albeit with a perturbation that depending on the chosen fermion to qubit mapping typically remains local in time but non-local in space (see \cref{fig:hubbard-circuit} (left) in \cref{appendix:circuit-decomposition}).
\begin{figure*}
\includegraphics{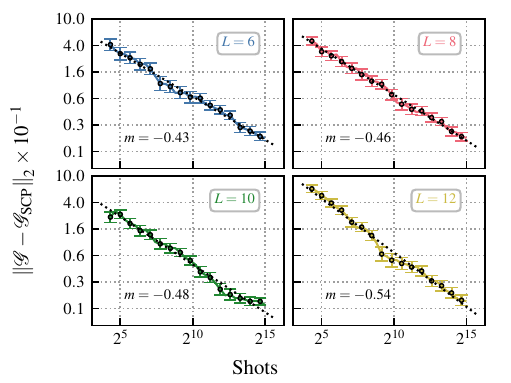}
\includegraphics{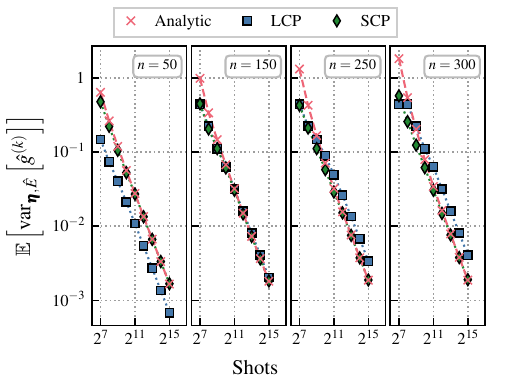}
\caption{\textbf{Convergence to ground truth and variance scaling.} We benchmark the \gls{lcp} and \gls{scp} algorithms on the one-dimensional Heisenberg model, analyzing the convergence to the exact dynamics and the scaling of the sampling variance with the number of circuit repetitions. \textit{Left}. Scaling with the number of shots of the norm of differences between the expected dynamics and \gls{scp} in log-log scale, for an increasing number of spins. As expected, convergence occurs with scaling equal to the inverse of the root of the samples, $\mathcal{O}\pt{1/\sqrt{\mathcal{S}}}$. The slopes $m$ of the fit lines are indicated in the lower left corners, and are in good qualitative agreement with the expected value of $-1/2$. \textit{Right}. Comparison of the mean of the variance estimates of the gradient in log-log scale of the \gls{lcp} and \gls{scp} methods as a function of the number of shots. The red curve shows \cref{eq:spsa-variance} for which the second term has been saturated setting $c_{\mathrm{SCP}}=1$ (see also \cref{appendix:variance}).}\label{fig:variance}
\end{figure*}
Due to the non-local nature of the operators involved, parallel measurements -- unlike in the bosonic case -- are not feasible in this scenario; however, the sampling advantage offered by \gls{scp} remains valid.

In the next section we present results for both these model Hamiltonians.

\section{\label{sec:results}Results}
\begin{figure*}
\includegraphics{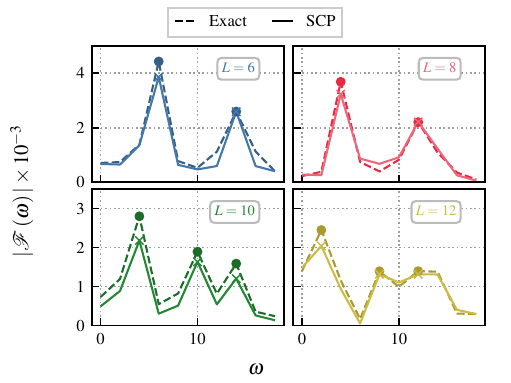}
\includegraphics{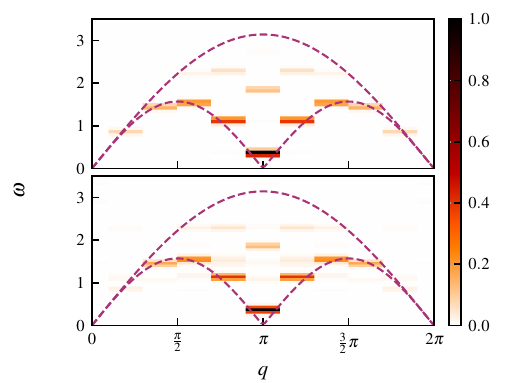}
\caption{\textbf{Fourier transform and dynamical structure factor.} We benchmark the \gls{scp} algorithm on the one-dimensional Heisenberg model and show the corresponding frequency-domain response and \gls{dsf} obtained from the \glspl{rgf}. \textit{Left}. Fourier transform of the exact and \gls{scp} curves for an increasing number of spins. \textit{Right}. Dynamical spin structure factor for a $10$-site spin chain using exact diagonalization and \gls{scp} (top and bottom panel respectively). The dashed magenta lines show the lower and upper limits of the two-spinon contributions as given in~\cite{ferrari_dynamical_2018}, excitations that carry most of the total intensity of the \gls{dsf}~\cite{karbach_two-spinon_1997}. The energies have been rescaled by a factor $2\pi/\max\pt{\omega_e}$; values have been normalized to $1$ for comparison, and the delta-functions in \cref{eq:dsf-final-expression} have been replaced by normalized Gaussians with $\sigma=0.2J$.}\label{fig:frequencies}
\end{figure*}
The simulations, where not otherwise specified, only considered statistical (or \emph{shot}) noise. We adopt natural units by setting the coupling and hopping parameter to $J^\alpha_r=J\equiv 1$. All energies are thus expressed in units of $J$, and time is measured in units of $1/J$. For all \gls{scp} simulations, we set $\varepsilon=0.1$.

To illustrate the practical use our proposed quantum algorithms, we show the calculation of $\mathcal{G}^{\alpha\beta}\pt{r,t}$ with $r=0$ for the Heisenberg model of \cref{sec:Heis} on $10$ spins for $\nperturb=100$ in \cref{fig:together} (left). The reported data, collected via numerical simulations, compare the exact dynamics obtained through direct matrix multiplication of the Trotter evolution with the results obtained from a simulation of the quantum circuits for \gls{lcp} and \gls{scp}.

We use a total budget of $\mathcal{S}=2^{14}$ shots: in the case of \gls{lcp}, this is divided equally across the $100$ data points reported in the plot. For \gls{scp}, instead, we distribute $\mathcal{S}$ across $P$ independent perturbations $\pg{\bm{\eta}_p}^{P-1}_{p=0}$ as described in \cref{eq:gradient-SPSA}. As shown in \cref{appendix:variance}, the variance of the $k$-th gradient component averaged over these perturbations when using $S$ shots each can be expressed as 
\begin{equation}\label{eq:spsa-variance}
    \var_{\bm{\eta}, \exfunc}\pq{\grad}=\frac{\var_{\bm{\eta}}\pq{g^{\pt{k}}}}{P}+\frac{c_{\mathrm{SCP}}}{PS\varepsilon^2}\;,
\end{equation}
where $c_{\mathrm{SCP}}$ is a constant. In \cref{eq:spsa-variance} the first term originates from the variability induced by the perturbations $\bm{\eta}_p$, while the second term represents the sampling noise associated with the operator averaging procedure required to reconstruct the output (i.e., the $\exfunc_s\pt{\pm\varepsilon\bm{\eta}_p}$ terms of \cref{eq:gradient-SPSA}) of each fixed circuit configuration. From a statistical point of view, the optimal shot allocation policy is then to choose $S=1$, i.e. to change $\pg{\bm{\eta}_p}^{P-1}_{p=0}$ at each shot. As already mentioned, this option may be quite demanding in practice, as it could lead to high circuit compilation and experimental control overheads in real devices unless, e.g. some form of parametric execution is available~\cite{fischer2024dynamical}. For this reason, we always assume a generic $S\geq 1$ in all our formal derivations, although we adopt the $S=1$ strategy in all numerical simulations. A numerical analysis of the $S>1$ case is reported in \cref{sec:resource-variance}.

Notice that both \gls{lcp} and \gls{scp} require two circuit evaluations per data point or gradient estimate. In the case of \gls{lcp}, as described in \cref{sec:lcp}, we reconstruct $F\pq{\pi/2}$ and $F\pq{-\pi/2}$ separately, using the full shot budget assigned to each individual time point for each evaluation. Similarly, for \gls{scp}, \cref{eq:gradient-SPSA} and \cref{eq:spsa-variance} assume that $\exfunc_s\pt{\varepsilon\bm{\eta}_p}$ and $\exfunc_s\pt{-\varepsilon\bm{\eta}_p}$ are estimated independently. In our implementation, we take $S=1$, meaning that each of the two required evaluations for a given perturbation is performed with a single shot. Thus, in both cases, the total number of circuit runs is $2\mathcal{S}$, with $\mathcal{S}$ independent perturbations available for \gls{scp}.

In \cref{fig:together} (right) we show a similar plot, this time for the time evolution of a $4$-site $1$D periodic Hubbard model, with $J=1$, $U=5$ (see \cref{eq:hubbard}) and $\nperturb=30$. On top of the comparison with the expected and Trotter dynamics for the \gls{scp} method, we also show the curves obtained by applying a depolarizing noise channel
\begin{equation}
    \mathcal{D}_\gamma\pt{\rho}=\pt{1-\gamma}\rho+\gamma\Tr\pq{\rho}\frac{\mathbb{1}}{2^\nqubits}\;,
\end{equation}
where $\gamma$ is the depolarizing error parameter, $0\leq\gamma\leq 4^\nqubits/\pt{4^\nqubits-1}$. We apply the channel to all two-qubit circuit gates for increasing values of $\gamma=5\cdot 10^{-3}$ and $\gamma=10^{-3}$, which are in the range of typical experiments. The top plot shows the (symmetric) Fourier transform of the signal: the relative error between the \gls{scp} spectra and the exact reference spectrum remains well below $1\%$ for each noise strength $\gamma$, confirming that \gls{scp} reproduces the relevant frequencies with high accuracy even in the presence of realistic depolarizing noise. \cref{fig:variance} (left) numerically confirms that the \gls{scp} estimates converge to the expected result with a scaling of the residual errors proportional to $1/\sqrt{\mathcal{S}}$. Furthermore, for local single-qubit operators such as the ones appearing in the Heisenberg chain, the errors do not significantly scale with system size.

The potential advantages of \gls{scp} over \gls{lcp} in terms of overall sampling costs are showcased in \cref{fig:variance} (right), where we report the average variance of the gradient's estimates for the Heisenberg model as a function of the total shot budget $\mathcal{S}$ and for a different number of points (and thus perturbations $\nperturb$) along the time axis, indicated by the figure in the top right corner. As the mesh becomes finer, one can easily see that the errors at fixed $\mathcal{S}$ increase for \gls{lcp}, where each additional point requires a separate experiment and consumes a definite fraction of the total shots $s_r=\lceil S/\nperturb\rceil$. Such increase is slower for \gls{scp}, with a crossover around $\nperturb\approx 150$. This suggests, as already stated before, that \gls{scp} might become particularly convenient in practical regimes of interest, where long simulation times and dense curve samplings are required to resolve higher frequencies. Since the observable is taken from the set $\pg{\sigma^\alpha_r}_{r\in\pq{0,\dots,\nqubits-1}}$, the variance for \gls{lcp} is given, according to its definition, by $\pt{1-\pa{\sigma^\alpha_r}}^2/N$. Notice that here the number of Trotter steps increases with the number of perturbations, even though the total simulation time is kept fixed for all curves. Our numerical estimates are confirmed by the analytic results obtained by a direct application of \cref{eq:spsa-variance}. For completeness, in \cref{tab:LCP-SCP-comparison} of \cref{sec:resource-variance} we show an overall methodological comparison of the two techniques.

\begin{figure*}
\includegraphics{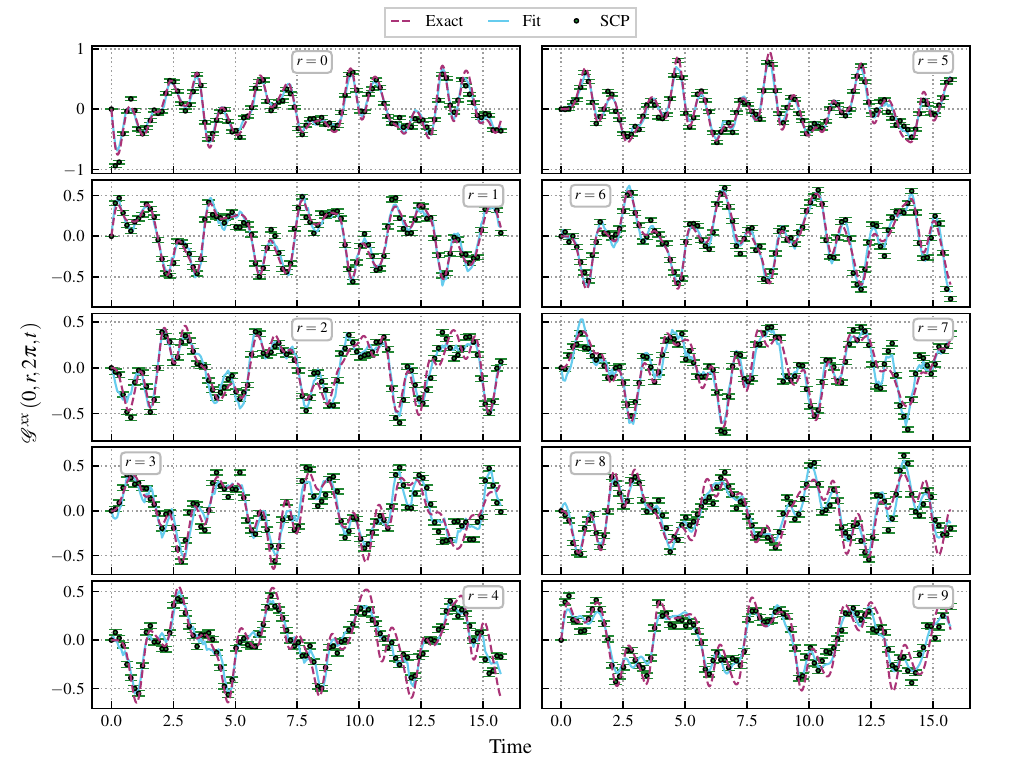}
\caption{\textbf{Green's functions and fits for the calculation of the dynamical structure factor.} Green's functions as in \cref{eq:green-spins} for a $10$-qubit chain for $r\in\pg{0,\dots,9}$, using $\mathcal{S}=2^{16}$ shots. The fits (cyan curves) are performed using \textit{iminuit}~\cite{iminuit}. Note that all the green data points are obtained by running a single circuit template.}\label{fig:group-picture}
\end{figure*}

Finally, we report in \cref{fig:frequencies} the frequency domain Fourier transforms of the generalized susceptibilities for Heisenberg chains of various lengths (left) and the calculation of the \gls{dsf} for the $10$-spin case (right), both computed via \gls{scp}. The \glspl{rgf} utilized for the calculation of the latter are depicted in \cref{fig:group-picture}. Each time trace is fitted using the Fourier form introduced in \cref{eq:fitting-function}, with the additional constraint that $\mathcal{G}^{\alpha\alpha}\pt{r,0}=0 \; \forall r$, as the equal time commutator $\pq{{\sigma^\alpha_r\pt{0}, \sigma^\alpha_{0}\pt{0}}}$ between any pair of local spin operators along the same axis vanishes. We note that the use of \gls{scp} allows us to retrieve all data points along the time traces -- which are required to carry out the temporal Fourier analysis and extract the $a_{r,e}^{\alpha\beta}$ coefficients and the frequencies $\omega_e$ -- with a single circuit instance equipped with random single-qubit perturbations. Furthermore, for any choice of $\alpha$, $\beta$ and $t$, all \glspl{rgf} $\pg{\mathcal{G}^{\alpha\beta}\pt{r,t}}_{r\in\pq{0,\dots,\nqubits-1}}$ can be computed in parallel by measuring commuting local Pauli operators. Within finite-size effects, our results are consistent with those obtained using variational Monte Carlo~\cite{ferrari_dynamical_2018}, restricted Boltzmann machines~\cite{hendry_chebyshev_2021} or a Bethe \emph{ansatz} approach~\cite{lake_multispinon_2013}.

\section{\label{sec:conclusions}Conclusions}

We presented a unified formulation that reduces the evaluation of \glspl{rgf} at discrete time points to the estimation of derivatives of a parametrized quantum-circuit output with respect to controllable perturbation parameters. By recasting \gls{lr} quantities as circuit derivatives, this framework clarifies that a broad range of circuit-differentiation tools -- deterministic and stochastic -- can be brought to bear on retarded response calculations. In this sense, the main outcome of this work is a general route to constructing families of \gls{rgf} estimators from a common circuit-perturbation representation; we illustrated this perspective with two representative instances, namely \gls{lcp} and \gls{scp}.

We validated the resulting protocols numerically on interacting spin and fermionic lattice models. Across the considered settings, the reconstructed dynamical correlations and derived spectral quantities (e.g., dynamical structure factors) are consistent with exact-diagonalization benchmarks, even under representative noise models.

Beyond near-term stochastic and deterministic estimators, the circuit-differentiation viewpoint also enables connections to more general estimation primitives. We provided a principled route to combining the present framework with gradient-estimation techniques in the fault-tolerant regime, potentially reducing sampling overheads and enabling a unified estimation of multiple response coefficients (e.g., a time grid of retarded correlators) within a single coherent procedure.

Several extensions are natural in light of this formulation. Our approach lends itself to incorporating recent developments that denoise and extend response functions by exploiting positive-definiteness constraints~\cite{kemper_positive_2023}. More broadly, the explicit mapping between response functions and circuit derivatives suggests additional theoretical and computational refinements, including estimator design, resource-aware compilation strategies, and measurement-efficient schemes for extracting multiple observables. Finally, the idea of accessing temporal properties in parallel may be further generalized using dedicated clock-qubit registers~\cite{diaz_parallel_2024}, offering another avenue to scale time-resolved quantum simulation beyond straightforward sequential sampling.
\begin{figure*}
    \centering
    \begin{quantikz}
        & \gate[2]{\sigma^x\sigma^x+\sigma^y\sigma^y+\sigma^z\sigma^z} &  \midstick[2,brackets=none]{=}
        & \ctrl{1} & \gate{R^x\pt{\gamma^x - \pi/2}} & \gate{H} & \ctrl{1} & \gate{H} & \ctrl{1} & \gate{R^x\pt{\pi/2}} & \\
        & & & \targ & & \gate{R^z\pt{\gamma^z}} & & \targ & & \gate{R^z\pt{-\gamma^y}} & \targ & & \gate{R^x\pt{-\pi/2}} &
    \end{quantikz}
    \caption{\textbf{Heisenberg circuit decomposition.} $3$-CNOT circuit decomposition for the digital quantum simulation of the $2$-qubit Heisenberg model. For each single-qubit rotation $R^\alpha\pt{\cdot}$, we define the angle as $\gamma^\alpha=2J^\alpha\tau$, where $\tau=T/\nperturb$ is the elementary time step, $T$ the total simulation time window and $\nperturb$ is the number of Trotter steps. The phase difference from the expected unitary is of $\me^{\ramuno\pi/4}$.}\label{fig:heisenberg-circuit}
\end{figure*}
\begin{figure*}
\begin{quantikz}
    \lstick{$q_A$} & \qw\gategroup[2,steps=5,style={dashed,rounded corners,inner xsep=1pt,inner ysep=10pt,draw=qviolet},background,label style={yshift=0.05cm,xshift=0cm, text=qviolet}]{{$\me^{-\ramuno\theta \sigma^z_A \sigma^x_B}$}} & \ctrl{1}\gategroup[2,steps=3,style={dashed,rounded corners,inner xsep=2pt,inner ysep=5pt,draw=qgray},background,label style={yshift=-0.95cm,xshift=0cm,text=qgray}]{{$\me^{-\ramuno\theta \sigma^z_A \sigma^z_B}$}}& \qw& \ctrl{1}& \qw & \qw \\
    \lstick{$q_B$} & \gate{R^y\pt{-\pi/2}}& \targ{}&\gate{R^z\pt{2\theta}}
    & \targ{} & \gate{R^y\pt{\pi/2}} \qw &
\end{quantikz}\hfill%
\begin{quantikz}
    & \gate{R^x\pt{\pi/2}} & \ctrl{1} & \gate{R^x\pt{2\theta}} & \ctrl{1} & \gate{R^x\pt{-\pi/2}} & \qw \\
    & \gate{R^x\pt{\pi/2}} & \targ{}  & \gate{R^z\pt{2\theta}} & \targ{} & \gate{R^x\pt{-\pi/2}} & \qw
\end{quantikz}
\caption{\textbf{Fermi-Hubbard circuit decomposition.} \textit{Left}. Circuit decomposition for implementing the exponential of $\sigma^z\sigma^z$ (gray rectangle) and $\sigma^z\sigma^x$ (purple rectangle). The case $\sigma^z\sigma^y$ is obtained by flipping the sign of the angles of both $R^y$ rotations and substituting $R^y\to R^x$. \textit{Right}. $2$-CNOT circuit decomposition of the $\sigma^x\sigma^x+\sigma^y\sigma^y$ exponential implementing the \gls{jw}-transformed hopping term (see \cref{eq:cre-ann-jw,eq:hubbard}).}\label{fig:hubbard-circuit}
\end{figure*}

\begin{acknowledgments}
We thank Julien Gacon and Matthis Lehmk\"uhler for useful discussions, as well as Anthony Gandon and Almudena Carrera Vazquez for valuable feedback on the manuscript. We would also like to thank the anonymous reviewers for their constructive comments and suggestions, which have helped to both improve the clarity and broaden the scope of this work. This research was supported by NCCR MARVEL, a National Center of Competence in Research, funded by the Swiss National Science Foundation (grant number 205602) and by RESQUE funded by the Swiss National Science Foundation (grant number 225229).
\end{acknowledgments}

\bibliographystyle{quantum}
\bibliography{bib}


\clearpage
\appendix
\section{\label{appendix:parameter-shift}Calculation of the retarded Green's function and parameter-shift rule}

In this section we show that the analytical form of the \gls{psr} in the case of a Pauli gate \emph{exactly} matches the one in \cref{eq:green}, thus establishing a link between the calculation of a derivative of a parametrized Pauli rotation and the calculation of the \gls{rgf}. We start by considering a parametrized gate in the form
\begin{equation}\label{eq:pauli-gate}
    U_\ell\pt{\theta_\ell}=\exp\pt{-\ramuno\frac{\theta_\ell}{2}P}\;,
\end{equation}
with $P$ a Pauli operator. Let us examine the action of \cref{eq:pauli-gate} in a parametrized quantum circuit, where the overall unitary transformation can be expressed as a product of $N$ unitaries,
\begin{equation}
    U\pt{\bm{\theta}}=\prod^{N-1}_{\ell=0}U_\ell\pt{\theta_\ell}\;.
\end{equation}
Without loss of generality -- since any Hermitian operator can be expanded as a linear combination of Pauli operators -- we may write a generic observable as $Q$, another Pauli operator. The action of the quantum circuit on $Q$ can be expressed by the function
\begin{equation}\label{eq:quantum-circuit-function}
    \mathcal{U}\pt{\bm{\theta}}=\Braket{\psi_{\ell-1} | U^\dagger_\ell\pt{\theta_\ell}Q_{\ell+1}U_\ell\pt{\theta_\ell} |\psi_{\ell-1}}\;,
\end{equation}
where we have absorbed any gate happening before the $\ell$-th gate in the initial state, $\ket{\psi_{\ell-1}}=\prod^{\ell-1}_{j=0}U_j\pt{\theta_j}\gs$ and any gate applied after in the observable $Q_{\ell+1} = \prod_{j=N-1}^{\ell+1} U_j^\dagger\pt{\theta_j} \, Q \, \prod_{j=\ell+1}^{N-1} U_j\pt{\theta_j}$. Since the function in \cref{eq:quantum-circuit-function} depends smoothly on the parameter $\theta_\ell$, it admits a well-defined gradient
\begin{equation}\label{eq:GF-link}
    \nabla_{\theta_\ell}\mathcal{U}\pt{\bm{\theta}}=-\frac{\ramuno}{2}\Braket{\psi_{\ell-1} | U^\dagger_\ell\pt{\theta_\ell}\pq{Q_{\ell+1}, P}U_\ell\pt{\theta_\ell} |\psi_{\ell-1}}\;,
\end{equation}
which follows from the derivative
\begin{equation}
    \nabla_{\theta_\ell}U_\ell\pt{\theta_\ell}=-\frac{\ramuno}{2}U_\ell\pt{\theta_\ell}P\;.
\end{equation}
For commutators involving Pauli operators, one can use the identity introduced in Ref.~\cite{mitarai_quantum_2018}
\begin{equation}
\pq{Q, P} = \ramuno\pt{U_\ell^\dagger \pt{\frac{\pi}{2}} Q U_\ell \pt{\frac{\pi}{2}} - U_\ell^\dagger \pt{-\frac{\pi}{2}} Q U_\ell \pt{-\frac{\pi}{2}} }\;,
\end{equation}
Substituting this into \cref{eq:GF-link} yields the \gls{psr},
\begin{equation}
    \nabla_{\theta_\ell}\mathcal{U}\pt{\bm{\theta}}=\frac{1}{2}\pq{\mathcal{U}\pt{\theta_\ell+\frac{\pi}{2}}-\mathcal{U}\pt{\theta_\ell-\frac{\pi}{2}}}\;.
\end{equation}
For our purposes, however, it is sufficient to stop at \cref{eq:GF-link}. Focusing on the first term of the commutator and ignoring for a moment the prefactor of $-\ramuno/2$, by expanding back the gates in the states and the observable we get
\begin{equation}
    \Braket{\psi_0 | \prod^{0}_{j=N-1}U^\dagger_j Q\prod^{N-1}_{j=\ell+1}U_j P U_\ell \prod^{\ell-1}_{j=0}U_j|\psi_0}\;,
\end{equation}
where we removed the $\theta$ dependency from the unitaries for ease of notation. We can now insert an identity $\prod^\ell_{j=0}U_j\prod^0_{j=\ell}U^\dagger_j$ before $P$, yielding
\begin{equation}\label{eq:psr-expansion}
    \Braket{\psi_0 | \prod^{0}_{j=N-1}U^\dagger_j Q\prod^{N-1}_{j=0}U_j\prod^0_{j=\ell}U^\dagger_jP\prod^\ell_{j=0}U_j|\psi_0}\;.
\end{equation}
With \cref{fig:fd-circuit} in mind, the unitaries can be identified as $N$ Trotter blocks with $\theta\equiv\tau=T/N$, where $T$ is the total evolution time. The Paulis in \cref{eq:psr-expansion} are thus evolved to times $Q\pt{T}$ and $P\pt{t^\prime}$ respectively, with $t^\prime=\ell\tau$. Since the second commutator term in \cref{eq:GF-link} produces the complex conjugate of \cref{eq:psr-expansion}, we finally come to
\begin{equation}
    \text{\cref{eq:GF-link}}=-\frac{\ramuno}{2}\Braket{\psi_0 | \pq{Q\pt{T},P\pt{t^\prime}} | \psi_0}\;.
\end{equation}
This expression corresponds to \cref{eq:green} up to a factor of $2$, which, as mentioned in the main text, takes into account the symmetries of the \glspl{rgf}. This final observation marks the connection between the \gls{psr} and the \gls{rgf}.

\section{\label{appendix:extension-gradient}Extension to gradient-estimation algorithms}

To make use of Theorem 4 in Ref.~\cite{huggins_nearly_2022} (from here on: T4), one needs to show that a function $f:\mathbb{R}^M\to\mathbb{R}$ is both analytic in its parameters and satisfies a bound on its higher-order partial derivatives of the form
\begin{align}
    &\bigl|\partial_\alpha f\pt{\bm{x}}\bigr|\le c^{k}\, k^{k/2}\;, & &c\in\mathbb{R}_+\;, k\in\mathbb{Z}_+
\end{align}
with respect to any collection of indices $\alpha\in\pg{0,\dots,M-1}^k$ and at the point of interest (in particular, at $\bm{x}=\bm{0}$). Below we verify these conditions for the functional introduced in \cref{eq:general-F}.  For clarity, $\partial_\alpha$ denotes a mixed partial derivative of total order $|\alpha|=k$. We discretize the source field $\mathrm{s}\pt{\bm{r}^\prime, t^\prime}$ on a bounded space-time grid and collect its values into a finite-dimensional parameter vector $\bm{\varepsilon}$, with components $\varepsilon_n\coloneq\mathrm{s}\pt{r_n, t_n}$. Under this discretization, the functional $\mathcal{F}\pq{\mathrm{s}}$ becomes a scalar-valued function of $\bm{\varepsilon}$. The dimension $M$ is the number of discretized space–time components of the source field, i.e., $M \equiv \dim\pt{\bm{\varepsilon}}$.\\
To match the probability-oracle framework of Ref.~\cite{huggins_nearly_2022}, we let $f\pt{\bm{\varepsilon}}=\frac{1+\mathcal{F}\pt{\bm{\varepsilon}}}{2}$ such that $\partial_\alpha f=\frac{\partial_\alpha\mathcal{F}\pt{\bm{\varepsilon}}}{2}$. We then let $\rho_0 = |\psi_0\rangle\langle\psi_0|$, so that $\|\rho_0\|_1 = 1$ (where $\|\cdot\|_1$ and $\|\cdot\|$ are the trace and operator norms respectively) and assume the observable is normalized, $\|o_{\bm{R}}\|=1$, without loss of generality, since any bounded observable can be rescaled and the corresponding normalization restored at the end.

Using H\"older's inequality and the unitary invariance of the operator norm,
\begin{align}
    |\mathcal{F}\pt{\bm{\varepsilon}}|&=|\Tr\pq{\rho_0 U^\dagger_{\bm{\varepsilon}} o_{\bm{R}} U_{\bm{\varepsilon}}}|\\ &\leq\|\rho_0\|_1\|U^\dagger_{\bm{\varepsilon}} o_{\bm{R}}U_{\bm{\varepsilon}}\|=\|o_{\bm{R}}\|\;,
\end{align}
so that $f\pt{\bm{\varepsilon}}\in\pq{0, 1}$, $\forall\bm{\varepsilon}$. For each $n$, we define the kick
\begin{equation}
    V_n\pt{\varepsilon_n}\coloneq\me^{-\ramuno\tau\varepsilon_n o_{\bm{r}}\pt{n\tau}}\;.
\end{equation}
The map
\begin{equation}
    \varepsilon_n\mapsto \me^{-\ramuno\tau\varepsilon_n o_{\bm{r}}\pt{n\tau}}
\end{equation}
is entirely analytic. Finite products of analytic matrix-valued functions are analytic; conjugation by unitaries preserves analyticity, as well as taking the expectation value, given that it is a linear operation. Thus, both $\mathcal{F}\pt{\bm{\varepsilon}}$ and $f\pt{\bm{\varepsilon}}$ are real-analytic on $\mathbb{R}^M$. For the bounds on higher-order partial derivatives it holds that
\begin{align}
    &\frac{\partial V_n\pt{\varepsilon_n}}{\partial\varepsilon_n}=-\ramuno\tau o_{\bm{r}}\pt{n\tau} V_n\pt{\varepsilon_n}\;, \\ 
    &\left\|\frac{\partial V_n\pt{\varepsilon_n}}{\partial\varepsilon_n}\right\|\le\tau \|o_{\bm{r}}\|\;.
\end{align}
Since each parameter $\varepsilon_n$ appears in exactly one factor of $U_{\bm{\varepsilon}}$, any $k$-th order partial derivative satisfies $\|\partial_\alpha U_{\bm{\varepsilon}}\|\le\pt{\tau\|o_{\bm{r}}\|}^k$ and similarly for $\partial_\alpha U^\dagger_{\bm{\varepsilon}}$. By deriving, one obtains
\begin{equation}
    \partial_\alpha\pt{U^\dagger_{\bm{\varepsilon}} o_{\bm{R}}U_{\bm{\varepsilon}}}=\sum_{\beta\subseteq\alpha}\pt{\partial_{\beta}U^\dagger_{\bm{\varepsilon}}}o_{\bm{R}}\pt{\partial_{\bar{\beta}}U_{\bm{\varepsilon}}}\;,
\end{equation}
i.e. each of the $k$ derivatives may act either on $U^\dagger_{\bm{\varepsilon}}$ or on $U_{\bm{\varepsilon}}$, yielding at most $2^k$ terms. Here, $\beta\subseteq\alpha$ denotes a sub-multi-index, with $\bar{\beta}$ its complement. Taking the operator norms,
\begin{align}
    \|\partial_\alpha\pt{U^\dagger_{\bm{\varepsilon}} o_{\bm{R}}U_{\bm{\varepsilon}}}\|&\leq\sum_\beta\|\partial_{\beta}U^\dagger_{\bm{\varepsilon}}\| \|o_{\bm{R}}\| \|\partial_{\bar{\beta}}U_{\bm{\varepsilon}}\|\\
    &\leq2^k\pt{\tau\|o_{\bm{r}}\|}^k\;.
\end{align}
For a bounded operator $X$, it holds $|\braket{\psi_0|X|\psi_0}|\leq X$, which leads to a bound on $|\partial_\alpha \mathcal{F}\pt{\bm{\varepsilon}}|$ and consequently on $|\partial_\alpha f\pt{\bm{\varepsilon}}|$,
\begin{equation}
    |\partial_\alpha f\pt{\bm{\varepsilon}}|\le 2^{k-1}\pt{\tau\|o_{\bm{r}}\|}^k\;.
\end{equation}
Finally, to match the requirements of T4 it is sufficient to set $c=2\tau\|o_{\bm{r}}\|$. We note that, since $c$ constitutes the upper limit of the precision of the quantum algorithm whose existence is guaranteed by T4, the number of Trotter steps $\nperturb$ indirectly affects its accuracy via $\tau$.

\section{\label{appendix:variance}Variance of the simultaneous circuit perturbation gradient}

The variance of each component $\grad$, with $k\in\pg{0,\dots,\nperturb-1}$ of the \gls{scp} gradient estimator defined in \cref{eq:gradient-SPSA} with respect to both shot noise and random circuit perturbations (that is, the variance of each of the $\nperturb$ elements in the $P$ instances) can be computed starting from the law of total variance,
\begin{align*}
    &\var_{\bm{\eta}, \exfunc}\pq{\grad}=\var_{\bm{\eta}}\pq{\eval_\efunc\pq{\grad | \bm{\eta}_0,\dots,\bm{\eta}_{p-1}}}+\\
    &+\eval_{\bm{\eta}}\pq{\var_\efunc\pq{\grad | \bm{\eta}_0,\dots,\bm{\eta}_{p-1}}}\;.
\end{align*}
Using the linearity of the expectation value, with $\eval\pq{\exfunc_s}\equiv\efunc_s$ and by introducing the notation $\exfunc^\pm_{s,p}=\exfunc_s\pt{\theta\pm\bm{\eta}_p\varepsilon}$, this can be rewritten as
\begin{align}
    &\var_{\bm{\eta}}\pq{\frac{1}{P}\sum^{P-1}_{p=0}\frac{\efunc\pt{\theta+\bm{\eta}_p\varepsilon}-\efunc\pt{\theta-\bm{\eta}_p\varepsilon}}{2\varepsilon}\eta^{\pt{k}}_p}+\label{eq:rewriting-total-variance-first}\\
    &+\eval_{\bm{\eta}}\pq{\frac{1}{\pt{PS2\varepsilon}^2}\var_\efunc\pq{\sum^{P-1}_{p=0}\sum^{S-1}_{s=0}\pt{\hat{E}^+_{s,p}-\hat{E}^-_{s,p}}\eta^{\pt{k}}_p}}\;.\label{eq:rewriting-total-variance-second}
\end{align}
Given that $\var_{\bm{\eta}}\pq{g^{\pt{k}}}$ is the same over i.i.d. different realizations of the perturbation vector $\bm{\eta}_p$, \cref{eq:rewriting-total-variance-first} can be expressed as $N^{-1}\var_{\bm{\eta}}\pq{g^{\pt{k}}}$. For \cref{eq:rewriting-total-variance-second}, we start by noticing that the sums of $\exfunc^+_s$ and $\exfunc^+_s$ are not correlated,
\begin{widetext}
\begin{align}
    &\text{\cref{eq:rewriting-total-variance-second}}=\frac{1}{\pt{PS2\varepsilon}^2}\eval_{\bm{\eta}}\pq{\sum^{P-1}_{p=0}\eta^{\pt{k}}_p\pq{\var_\efunc\pt{\sum^{S-1}_{s=0}\exfunc^+_{s,p}}+\var_\efunc\pt{\sum^{S-1}_{s=0}\exfunc^-_{s,p}}}}=\\
    =&\frac{1}{\pt{PS2\varepsilon}^2}\eval_{\bm{\eta}}\pq{\sum^{P-1}_{p=0}\eta^{\pt{k}}_pS\pq{\nu^2\pt{\theta+\bm{\eta}_p\varepsilon}+\nu^2\pt{\theta-\bm{\eta}_p\varepsilon}}}=\frac{1}{PS\varepsilon^2}\eval_{\bm{\eta}}\pq{\frac{\nu^2\pt{\theta+\bm{\eta}_0\varepsilon}+\nu^2\pt{\theta-\bm{\eta}_0\varepsilon}}{4}\eta_0^{\pt{k}}}\label{eq:second-term}\;,
\end{align}
\end{widetext}
where in the two equalities of \cref{eq:second-term} we first denote by $\nu^2$ the variance of the function computed at a point and then we fix the realization of $\eta^{\pt{k}}_p$ to $\eta^{\pt{k}}_0$. We can identify for brevity the expectation value in \cref{eq:second-term} as
\begin{equation}
    c_{\mathrm{SCP}}\equiv\eval_{\bm{\eta}}\pq{\frac{\nu^2\pt{\theta+\bm{\eta}_0\varepsilon}+\nu^2\pt{\theta-\bm{\eta}_0\varepsilon}}{4}\eta_0
    ^{\pt{k}}}\;.
\end{equation}
In this case the expectation value cannot be distributed to the terms $\nu^2$ and $\eta^{\pt{k}}_0$ separately since they are not independent variables. Note that $c_{\mathrm{SCP}}$ is upper-bounded by $1/2$; in fact, $\nu^2=\pt{1-\pa{\sigma^\alpha_r}}^2/N\leq 1$, since (as for the \gls{lcp} case explained in \cref{sec:results}) only one Pauli is evaluated for each parameter extraction and thus, in particular, for $\theta\pm\bm{\eta}_0\varepsilon$.

We can piece this together, obtaining for the variance the final expression
\begin{equation}
    \var_{\bm{\eta}, \exfunc}\pq{\grad}=\frac{\var_{\bm{\eta}}\pq{g^{\pt{k}}}}{P}+\frac{c_{\mathrm{SCP}}}{PS\varepsilon^2}\;.\label{eq:final-variance-expression}
\end{equation}
From here, one can expand the variance appearing in the first term to first order as
\begin{widetext}
\begin{align}
    &\var_{\bm{\eta}}\pq{g^{\pt{k}}}=\var_{\bm{\eta}}\pq{\frac{\efunc\pt{\theta+\bm{\eta}_0\varepsilon}-\efunc\pt{\theta-\bm{\eta}_0\varepsilon}}{2\varepsilon}\eta^{\pt{k}}_0}=\\
    &=\var_{\bm{\eta}}\pq{\bm{\eta}^\transpose_0\nabla\efunc\pt{\theta}\eta^{\pt{k}}_0}+\mathcal{O}\pt{\varepsilon^2}=\var_{\bm{\eta}}\pq{\sum^{d-1}_{n=0}\frac{\partial}{\partial\theta_n}\efunc\pt{\theta}\eta^{\pt{n}}_0\eta^{\pt{k}}_0}+\mathcal{O}\pt{\varepsilon^2},
    =\label{eq:taylor-expansion}\\
    &=\pt{\textstyle\sum_{nm}\partial_n E\pt{\theta}\partial_{m}\efunc\pt{\theta}\eval_{\bm{\eta}}\pq{\eta^{\pt{n}}_0\eta^{\pt{m}}_0}}-\pt{\textstyle\sum_n\partial_n\efunc\pt{\theta}\eval_{\bm{\eta}}\pq{\eta^{\pt{n}}_0\eta^{\pt{k}}_0}}^2+\mathcal{O}\pt{\varepsilon^2}=\\
    &=\textstyle\sum_n\pt{\partial_n\efunc\pt{\theta}}^2-\pt{\partial_k\efunc\pt{\theta}}^2+\mathcal{O}\pt{\varepsilon^2}=\lVert\nabla\efunc\rVert^2_2-\pt{\nabla\efunc^{\pt{k}}}^2+\mathcal{O}\pt{\varepsilon^2}\;,\label{eq:result-first-term}
\end{align}
\end{widetext}
where the $\mathcal{O}\pt{\varepsilon^2}$ term in the first equivalence of \cref{eq:taylor-expansion} is given by a first-order Taylor expansion of the terms $\efunc\pt{\theta\pm\bm{\eta}_0\varepsilon}$ and is precisely the bias in the \gls{spsa} gradient estimator $\eval\pq{\grad}=\nabla E^{\pt{k}}$. In \cref{eq:result-first-term} we make use of $\eval_{\bm{\eta}}\pq{\eta^{\pt{n}}_0\eta^{\pt{m}}_0}=\delta_{nm}$.

By expressing \cref{eq:final-variance-expression} in terms of the total shot budget $\mathcal{S} = P S$, one observes that while the second term depends jointly on $\mathcal{S}$ only (which we assume fixed), the first becomes linear in $S$. This implies that the optimal shot distribution strategy is to use a single shot ($S = 1$) per randomization and maximize the number of independent samples $P$. With this allocation, we also recover the expected asymptotic scaling noted in \cref{sec:results} of $\var_{\bm{\eta}}\pq{g^{\pt{k}}} \propto \mathcal{O}\pt{1/\sqrt{\mathcal{S}}}$.
\begin{figure}
\includegraphics{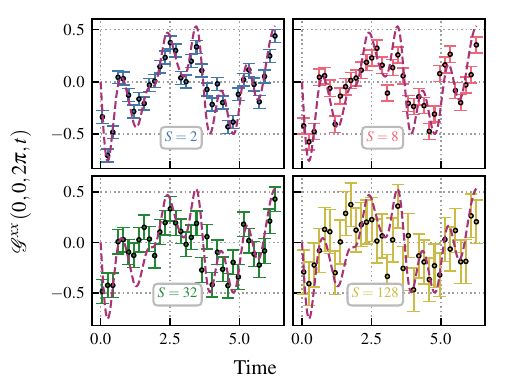}
\caption{\textbf{Green's function using \gls{scp} for \bm{$S\neq 1$}.} We compare the observable shown in \cref{fig:together} (left) for a (suboptimal) choice $S>1$ of the number of randomization shots. The panels, ordered left to right and top to bottom, correspond to increasing values of $S$, while the magenta dashed curve denotes the reference dynamics.}\label{fig:non-optimal-s}
\end{figure}
\section{\label{sec:resource-variance}Resource and variance comparison of the local- and simultaneous circuit perturbation protocols}

As mentioned in \cref{sec:SCP} and analyzed in \cref{appendix:variance}, the optimal shot strategy distribution is obtained by using $S=1$, i.e., a single shot per randomization of the perturbation vector. To analyze this trade-off further, \cref{fig:non-optimal-s} illustrates the behavior of an $xx$ \gls{rgf} under the \gls{scp} protocol as the number of randomization shots $S$ is increased above the optimal value, for a fixed total shot budget. Increasing $S$ leads to a systematic amplification of the variance, degrading the reconstruction relative to the reference dynamics. \cref{tab:LCP-SCP-comparison} complements this analysis by summarizing the resource requirements and shot-noise scaling of the \gls{lcp} and \gls{scp} protocols. While \gls{lcp} requires $N$ independent circuit executions with $S$ shots each, \gls{scp} executes a single circuit template with total shot budget $\mathcal{S}=PS$, whose variance follows the expression in \cref{eq:spsa-variance}.

In addition to statistical fluctuations, one should also consider Trotterization errors arising from the digital simulation of the dynamics. As discussed in \cref{appendix:parameter-shift}, the mapping between circuit derivatives and dynamical correlation functions is exact, so that both \gls{lcp} and \gls{scp} only inherit the Trotter error already present in the underlying time evolution. In \gls{lcp}, circuits targeting shorter evolution times can be chosen to be correspondingly shallower, whereas \gls{scp} always accumulates the Trotter error associated with the maximal evolution time. However, in the regime relevant for our benchmarks, where the Trotter step is chosen small enough to faithfully reproduce the dynamics up to the final time, our numerics indicate that shot noise remains the dominant source of error compared to Trotterization.
\begin{table}
\setlength{\tabcolsep}{4pt}
\renewcommand{\arraystretch}{1.5}
\centering
    \begin{tabular}{ccc}
    \hline
    \multicolumn{3}{c}{\textbf{Circuit resources}}\\
    \hline
    Protocol & Depth & Circuit templates \\
    \hline
    \gls{lcp} & $D_{U\pt{T}}+D_{\text{pert}}$ & $P$ \\
    \gls{scp} & $D_{U\pt{T}}+PD_{\text{pert}}$ & $1$ \\
    \hline
    \hline
    \multicolumn{3}{c}{\textbf{Sampling resources}}\\
    \hline
    Protocol & Shots/circuit & Variance per point \\
    \hline
    \gls{lcp} & $S$ & $1/S$ \\
    \gls{scp} & $PS$ & \cref{eq:spsa-variance} \\
    \hline
\end{tabular}
\caption{\textbf{Comparison of the \gls{lcp} and \gls{scp} protocols.} The table reports the circuit depth required to evaluate the \gls{rgf} over a time interval $\pq{0, T}$, the number of circuit executions, the number of shots per execution and the corresponding variance due to shot noise at each time instance for a total shot budget $\mathcal{S}=PS$. We denote the depth of the time evolution circuit and of the single circuit perturbation by $D_{U\pt{T}}$ and $D_{\text{pert}}$, respectively.}\label{tab:LCP-SCP-comparison}
\end{table}

\section{\label{sec:dsf-calculation}Dynamical structure factor calculation}

To compute \cref{eq:dsf-final-expression} one possible approach is to fit the curves of the \glspl{rgf} obtained via \gls{scp} and extract the frequencies $\omega_e$. To this end, we first obtain an ordered list of candidate frequencies by performing a Fourier transform on the observed data and extracting the most dominant peaks in the power spectrum. We then construct a sequence of models, each incorporating one additional frequency from this list, moving from a single-frequency model up to a predefined maximum. For each model, we employ a nonlinear least-squares minimization to fit the sine parameters (amplitudes and frequencies) appearing in \cref{eq:fitting-function}, minimizing the overall chi-squared statistic. After fitting, we calculate the \gls{bic}~\cite{BIC} for each model to balance goodness-of-fit against model complexity. The final model is chosen as the one yielding the lowest \gls{bic}, thereby retaining only those frequencies that produce a statistically robust improvement in fit without excessively increasing the parameter count.

We emphasize that this is only one possible way to extract frequencies, which is not the main focus of this paper, and that more sophisticated methods exist in principle. The technique described above, with the combination of a Fourier-based warm start and iterative fitting, provides both a reasonable initialization and a systematic mechanism for pattern selection.

\section{\label{appendix:circuit-decomposition}Circuit decomposition for the Heisenberg and Fermi-Hubbard model}

The Suzuki-Trotter expansion of the Heisenberg Hamiltonian in \cref{eq:heisenberg} breaks the evolution into two-body blocks that can be applied in parallel on even and odd qubits. Each of these blocks, shown in \cref{fig:heisenberg-circuit}, can be implemented with a 3-CNOT circuit~\cite{vidal_universal_2004}. The circuit perturbations, as mentioned in the main text, are in this case just one-qubit rotations implemented by single-qubit gates $R^\alpha\pt{\cdot}$, with the axis of the rotation pointing in the direction of the applied perturbation.

The quantum simulation of the Fermi-Hubbard model requires the implementation of both the hopping and interaction terms in \cref{eq:hubbard}. After applying the \gls{jw} transformation, the interaction term leads to products of single and double Pauli $\sigma^z$ operators. The exponentials of these terms can be implemented through single-qubit $z$-rotations and, in the case of two-body terms, through the circuit shown in \cref{fig:hubbard-circuit} (left, gray rectangle). The hopping term, on the other hand, gives rise - according to \cref{eq:cre-ann-jw} - to Pauli strings of the form $\mathcal{Z}\sigma^\alpha$, with $\alpha = x, y$. Knowing how to build exponentials of $\sigma^z$ strings, these terms can be realized in the circuit by applying suitable basis transformations through single-qubit rotations. Specifically, the following relations hold:
\begin{align}
    &R^y\pt{\pi/2}\sigma^zR^y\pt{-\pi/2} = \sigma^x\;,\\
    &R^x\pt{-\pi/2}\sigma^zR^x\pt{\pi/2} = \sigma^y\;.
\end{align}
An example of this procedure is illustrated in \cref{fig:hubbard-circuit} (left, purple rectangle), where the exponential of a $\sigma^z\sigma^x$ term is implemented. This provides a representative case of a circuit perturbation in the fermionic setting, where, in the general case of correlations between distant indices, chains of CNOT gates span between the first and last qubits involved (denoted respectively $q_A$ and $q_B$), rendering the operation non-local in space while remaining local in time. Circuits implementing perturbations of the form $\sigma^{x/y}$, followed by measurements of the corresponding observable conjugated by the parity operator $\mathcal{P}$ (see main text), yield an estimate of the \gls{rgf} according to either the \gls{lcp} or \gls{scp} approach. Finally, \cref{fig:hubbard-circuit} (right) shows the two-CNOT decomposition of the full hopping term $\pt{\sigma^x \sigma^x + \sigma^y \sigma^y}$.

\end{document}